\documentclass[preprint2]{aastex}
\usepackage{CJK}
\usepackage{enumitem}
\usepackage{amsmath} 
\newcommand{\angstrom}{\textup{\AA}}
\newcommand{\Oline}[1]{\textrm{[O{\footnotesize{#1}}]}} 
\newcommand{\Ne}[1]{\textrm{[Ne{\footnotesize{#1}}]}}

\newcommand{\N}[1]{$N_\textrm{\footnotesize{H}}$}
\newcommand{\Mg}[1]{\textrm{Mg{\footnotesize{#1}}}}
\newcommand{\Ca}[1]{\textrm{Ca{\footnotesize{#1}}}}

\newcommand{\C}[1]{\textrm{C{\footnotesize{#1}}}}
\newcommand{\lbd}[1]{$\lambda_{\textrm{Edd}}$}
\newcommand{\mbh}[1]{$M_{\textrm{BH}}$}
\shorttitle{X-ray Obscured Type 1 AGN}
\shortauthors{Liu et al.}
\begin{document}
\begin{CJK*}{UTF8}{gkai}

\title{Probing AGN Inner Structure with X-ray Obscured Type 1 AGN}

\author{Teng Liu（刘腾） \altaffilmark{1}, Andrea Merloni \altaffilmark{1}, Jun-Xian Wang（王俊贤） \altaffilmark{2,3}, Paolo Tozzi \altaffilmark{4}, Yue Shen \altaffilmark{5,6}, Marcella Brusa \altaffilmark{7,8}, Mara Salvato \altaffilmark{1}, Kirpal Nandra \altaffilmark{1}, Johan Comparat \altaffilmark{1}, Zhu Liu \altaffilmark{9}, Gabriele Ponti \altaffilmark{1}, Damien Coffey \altaffilmark{1}}
\altaffiltext{1}{Max-Planck-Institut f\"ur extraterrestrische Physik, Giessenbachstrasse 1, D-85748 Garching bei M\"unchen, Germany}
\altaffiltext{2}{CAS Key Laboratory for Research in Galaxies and Cosmology, Department of Astronomy, University of Science and Technology of China, Hefei 230026, China}
\altaffiltext{3}{School of Astronomy and Space Science, University of Science and Technology of China, Hefei 230026, China}
\altaffiltext{4}{Istituto Nazionale di Astrofisica (INAF) -- Osservatorio Astrofisico di Firenze, Largo Enrico Fermi 5, I-50125 Firenze}
\altaffiltext{5}{Department of Astronomy, University of Illinois at Urbana-Champaign, Urbana, IL 61801, USA}
\altaffiltext{6}{National Center for Supercomputing Applications, University of Illinois at Urbana-Champaign, Urbana, IL 61801, USA}
\altaffiltext{7}{Dipartimento di Fisica e Astronomia, Universit\'a di Bologna, viale Berti Pichat 6/2, I-40127 Bologna, Italy}
\altaffiltext{8}{INAF -- Osservatorio Astronomico di Bologna, via Ranzani 1, I-40127 Bologna, Italy}
\altaffiltext{9}{Key Laboratory of Space Astronomy and Technology, National Astronomical Observatories (NAOC), Chinese Academy of Sciences, Datun Rd 20A, Beijing 100012 China}

\begin{abstract}
Using the X-ray-selected active galactic nuclei (AGN) from the XMM-XXL north survey and the SDSS Baryon Oscillation Spectroscopic Survey (BOSS) spectroscopic follow-up of them,
we compare the properties of X-ray unobscured and obscured broad-line AGN (BLAGN1 and BLAGN2; \N{H} below and above $10^{21.5}$~cm$^{-2}$), including their X-ray luminosity $L_X$, black hole mass, Eddington ratio \lbd{}, optical continuum and line features.
We find that BLAGN2 have systematically larger broad line widths and hence apparently higher (lower) \mbh{} (\lbd{}) than BLAGN1.
We also find that the X-ray obscuration in BLAGN tends to coincide with optical dust extinction, which is optically thinner than that in narrow-line AGN (NLAGN) and likely partial-covering to the broad line region.
All the results can be explained in the framework of a multi-component, clumpy torus model by interpreting BLAGN2 as an intermediate type between BLAGN1 and NLAGN in terms of an intermediate inclination angle.
\end{abstract}


\section{Introduction}
Within the basic scheme of AGN unification model, both the differences between X-ray unobscured (X-ray type 1) and obscured (X-ray type 2) AGN and between broad-line (optical type 1) and narrow-line (optical type 2) AGN are determined by inclination angles with respect to an obscuring dusty ``torus''.
This axisymmetric ``torus'' plays an essential role in the unification model \citep{Antonucci1993}.
However, even recent ALMA high-resolution observations could only resolve a rotating circumnuclear disk for the nearby Seyfert galaxy NGC 1068 \citep{GarciaBurillo2016,Imanishi2018}.
The detailed structure of the ``torus'' is unclear, not to mention the physical mechanism that regulates it.
Especially, for a portion of AGN, the optical and X-ray classifications of type 1 and type 2 disagree with each other, complicating the understanding of the ``torus'' \citep{Brusa2003,Perola2004,Merloni2014,Davies2015}.

A large amount of work has been devoted to the study of the correlation between X-ray obscuration and luminosity.
Generally, relatively higher column densities (or larger obscured fractions) are found at lower luminosities \citep[e.g.,][]{1982Lawrence,2006Treister,Hasinger2008,Brightman2011a,2011Burlon,Lusso2013,2014Brightman}.
However, until reliable black hole masses (\mbh{}) for a sample of AGN are measured accurately, one could not clearly reveal the correlation between the obscuration and the \mbh{}-normalized accretion rate (i.e., the Eddington ratio, \lbd{}), which is considered as the main physical driver of the principle component of AGN properties \citep[e.g.,][]{Boroson1992,Sulentic2000}.
Using the Swift-BAT selected local AGN sample, \citet{Ricci2017} found that the AGN obscured fraction is mainly determined by the \lbd{} rather than the luminosity, and concluded that the main physical driver of the torus diversity is \lbd{}, which regulates the torus covering factor by means of radiation pressure.
To test the role of \lbd{} in regulating the obscuration of AGN, the \mbh{} of X-ray obscured and unobscured AGN must be measured consistently to avoid possible biases.
Except for the tens of AGN whose \mbh{} could be measured by reverberation mapping or dynamical methods \citep[e.g.,][]{Peterson2004}, generally, the \mbh{} of the X-ray unobscured AGN are measured on the basis of the broad line widths and the continuum luminosity (single-epoch virial method); for X-ray obscured AGN, the \mbh{} are inferred on the basis of the empirical relation between the \mbh{} and stellar velocity dispersion \citep[e.g.,][]{Ho2012,Bisogni2017b,Koss2017}.
The X-ray obscuration presented in a small fraction of broad-line AGN (BLAGN) provides a great tool for this test, since the \mbh{} of the X-ray unobscured and obscured BLAGN (BLAGN1 and BLAGN2) can be measured consistently using the same method.
Even then, the single-epoch \mbh{} must be used with caution, considering that the virial $f$ factor can be inclination dependent \citep{Wills1986,Risaliti2011,Pancoast2014,Shen2014,Bisogni2017,MejiaRestrepo2018}.

In BLAGN, whose broad line region (BLR) is visible, it is unclear what causes the X-ray obscuration.
The X-ray obscuring material might be dust-free and therefore transparent to optical emission from the accretion disc and BLR \citep{Merloni2014,Davies2015,Liu2016}; or it might be a dusty cloud blocking only the central engine (accretion disc and corona) but not the BLR because of geometric reasons, e.g., small obscuring cloud moving across the line of sight of the X-ray emitting corona \citep[e.g.,][]{Risaliti2002,Maiolino2010}.
Study of multi-band emission and obscuration of BLAGN2 could reveal rich information about the AGN environment close to the black hole.

In this work, we study the BLAGN in the XMM-XXL north survey.
We introduce the data in \S~2, investigate the X-ray obscuration of BLAGN in \S~3 and the optical spectral properties of them in \S~4.
The results are summarized and discussed in \S~5.

\section{The Data}
\label{sec:sample}
\subsection{The XXL-BOSS BLAGN Sample}
The XMM-XXL survey provide a large catalog (8445) of point-like X-ray sources.
3042 of them with R\footnote{Throughout the paper, R band corresponds to the SDSS observed band.} band AB magnitude between $15.0$ and $22.5$ were followed up by the BOSS spectrograph \citep{Georgakakis2011,Menzel2016}.
Based on the widths of the optical emission lines, i.e., H$\beta$, \Mg{II}, or \C{IV}, \citet{Liu2016}  measured the \mbh{} of the BLAGN in the catalog.
For sources with reliable redshift measurement and optical classification, \citet{Liu2016} used a Bayesian method \citep{Buchner2014} to measure the \N{H} and rest-frame 2-10~keV luminosity $L_X$.
To select the X-ray obscured sources, we use the same divide at \N{H}$=10^{21.5}$~cm$^{-2}$ as used in \citet{Merloni2014}, who found that this value provides the most consistent X-ray and optical classifications.
Among the XXL BLAGN, $>20\%$ of them have \N{H}$>10^{21.5}$~cm$^{-2}$, and if only sources with $>50$ net counts are considered, the fraction is $\sim10\%$.

With respect to the \mbh{} measured on the basis of hydrogen Balmer lines, \mbh{} measured using \Mg{II} is broadly consistent, while \C{IV}-based measurement can be systematically biased \citep{Shen2008,Shen2012,Coatman2017}.
Therefore, we select only the sources with \mbh{} measured using H$\beta$ and \Mg{II}.
This is roughly equivalent to excluding the sources at $z\gtrsim2.5$ whose \Mg{II} line is out of the BOSS wavelength range (3600--10000$\angstrom$).
When having both H$\beta$ and \Mg{II} measurements of \mbh{}, we choose the one with smaller uncertainty.

\begin{figure}[htbp]
\epsscale{1}
\plotone{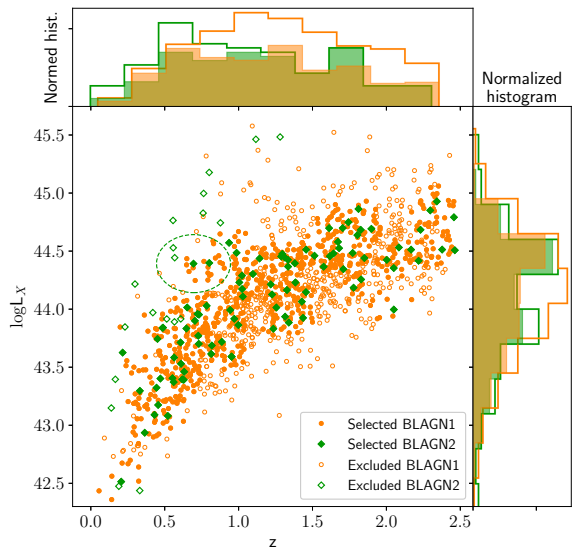}
\caption{The X-ray luminosity -- redshift scatter plot of the BLAGN1 (orange) and BLAGN2 (green) in sample ``0''.
In the ``same $L_X$--$z$'' re-selection (see \S~\ref{sec:comparison} for details), the solid points are selected and the empty ones are excluded.
The maximum distance of one point is shown with a dashed circle as an example.
Normalized histograms of luminosity and redshift are also shown for BLAGN2 (green) and BLAGN1 (orange) samples, in which the empty part corresponds to the excluded sources.
}
\label{fig:reselect}
\end{figure}

Based on the X-ray spectral analysis presented in \citet{Liu2016}, we classify sources with $\log$\N{H}$>21.5$ and with 1$\sigma$ lower limit of $\log$\N{H} above $21$ as BLAGN2, and those with $\log$\N{H}$\leqslant 21.5$ as BLAGN1.
We exclude sources with low optical S/N (\texttt{SN\_MEDIAN\_ALL} $<1.6$, see Appendix B of \citet{Menzel2016}) and a few low X-ray S/N sources whose intrinsic 2-10~keV luminosities are not well constrained (the width of $\log L_X$ 1$\sigma$ confidence interval $>0.5$). 
We also exclude a few sources whose broad lines are very weak through visual inspection, because they might be actually narrow-line AGN (NLAGN) with false detection of broad lines.
By now, our sample comprises 1172 BLAGN1 and 113 BLAGN2, whose luminosity--redshift distributions are shown in the central panel of Fig.~\ref{fig:reselect}.
A code name ``0'' is assigned to this sample.
However, this is not yet the eventual sample.
Thanks to our analysis of the source properties in the following sections, we notice that it is best to further exclude a few sources whose nature is uncertain (highly obscured or having a very-low accretion rate).
These additional filters give rise to the eventual sample ``1'', see \S~\ref{sec:comparison} for details.

\subsection{The BOSS Spectra}
\label{Sec:specslope}

\begin{figure}[htbp]
\epsscale{1}
\plotone{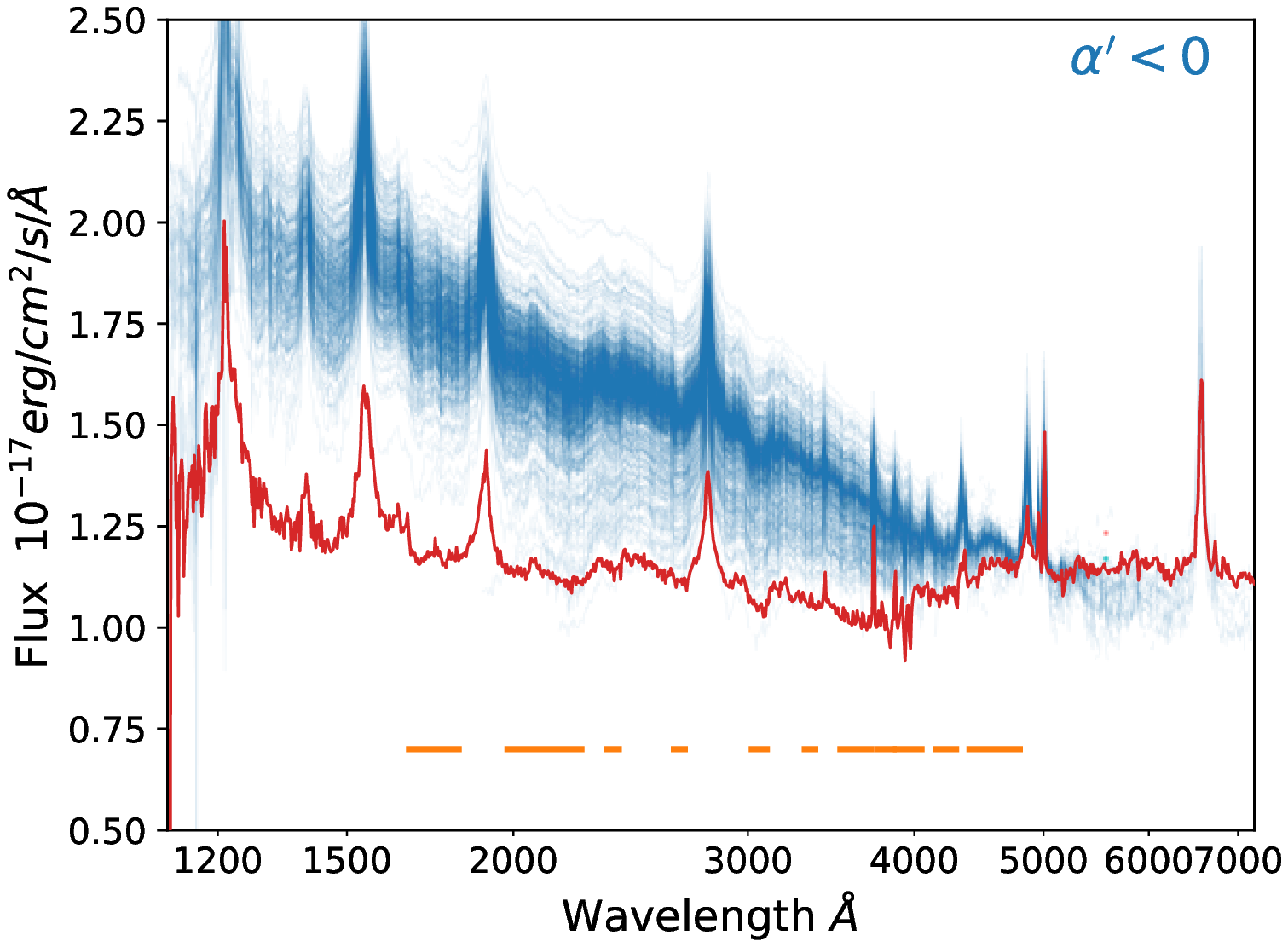}
\plotone{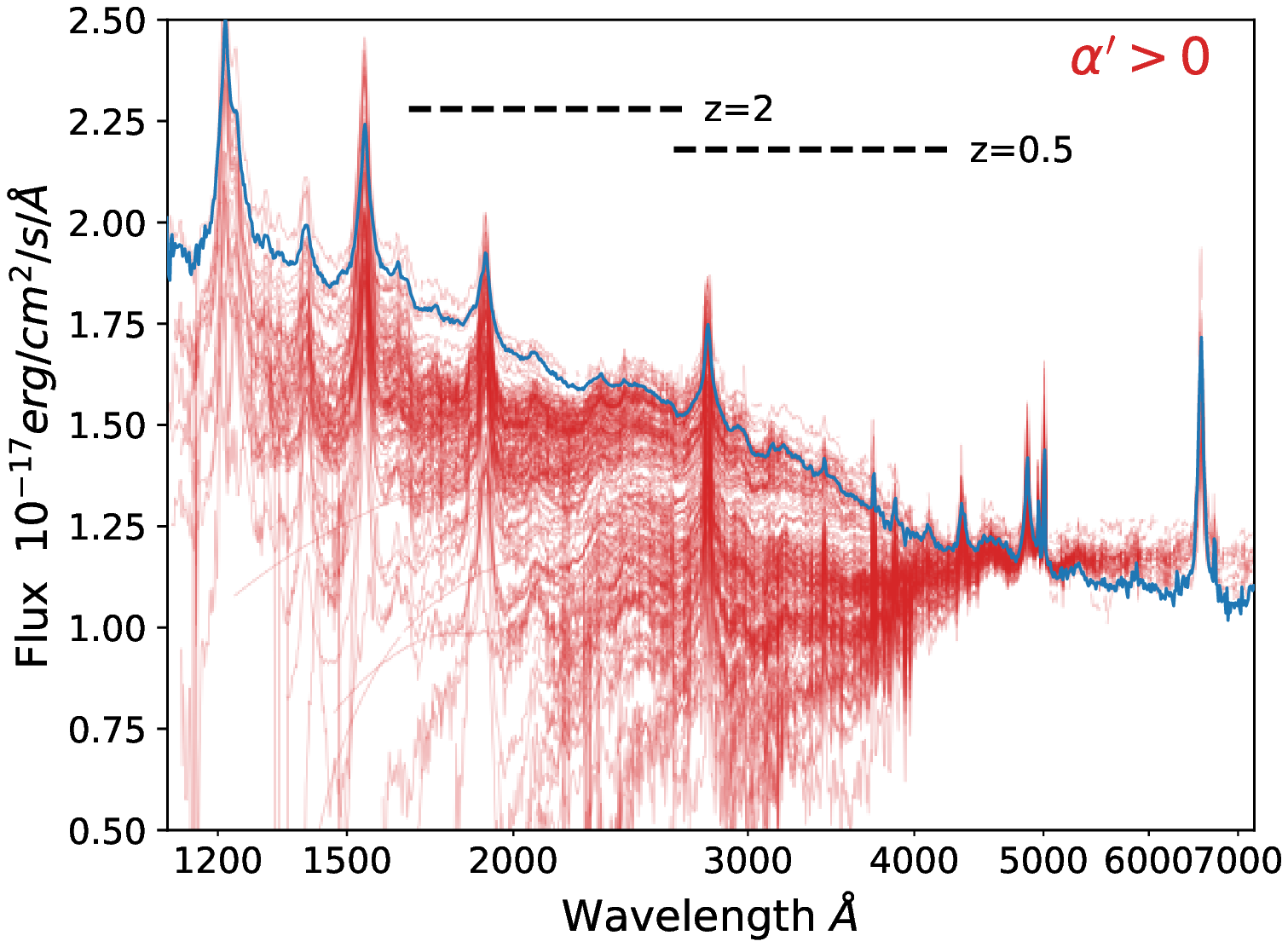}
\plotone{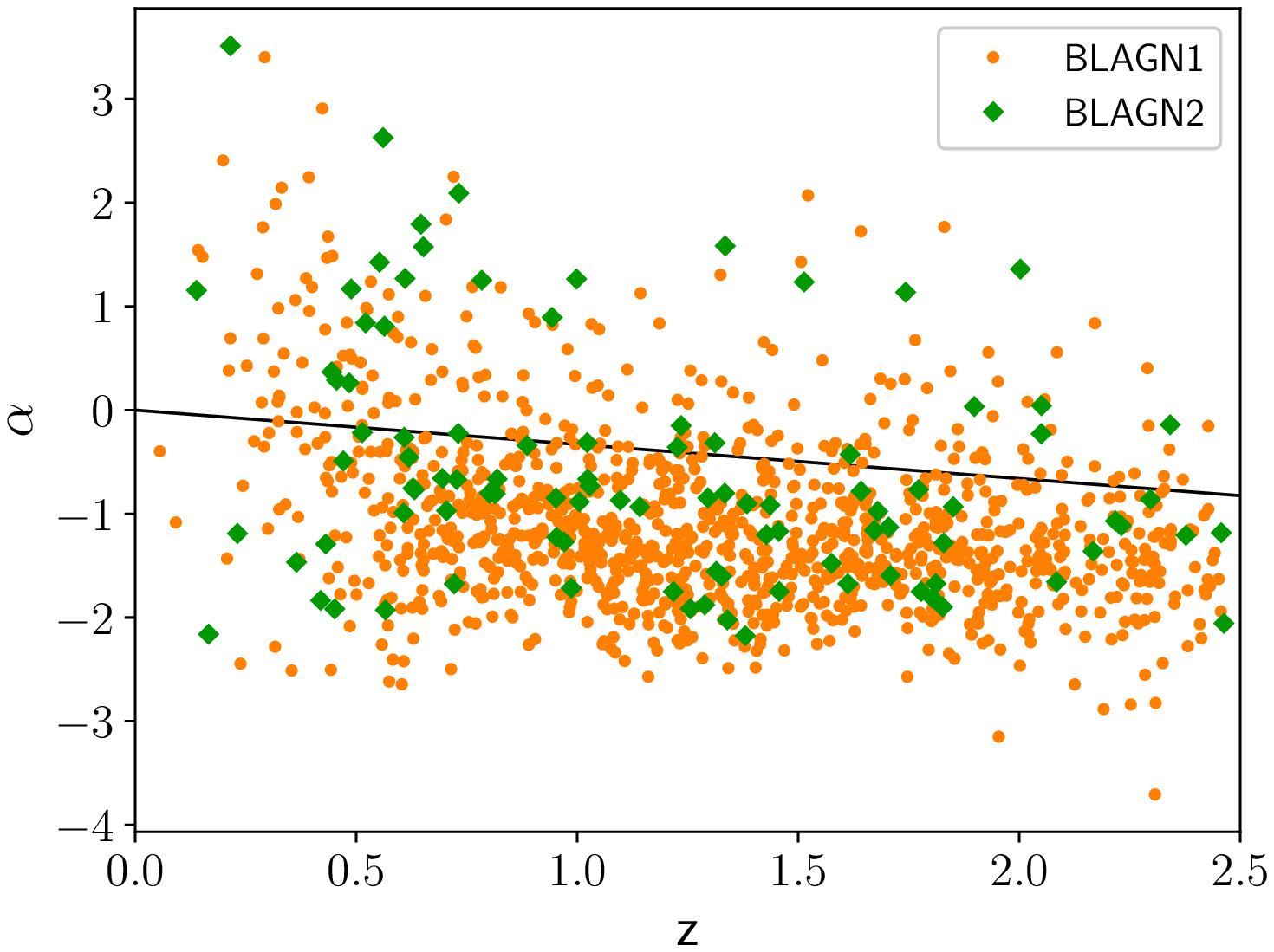}
\caption{
\footnotesize{
The blue lines in the top panel and the red lines in the middle panel show the galactic-extinction-corrected, rest-frame spectrum of each source for the blue ($\alpha'<0$) and red ($\alpha'>0$) BLAGN, respectively.
All the spectra are normalized to the median composite spectrum of the blue BLAGN, which is plotted as the blue line in the middle panel, and the median composite spectrum of the red BLAGN is plotted as the red line in the top panel (see \S~\ref{sec:stackingmethod} for details about the composite spectra).
We plot the best-fit models from the SDSS pipeline instead of the real data in order to have a clear look even in the low S/N cases. 
The ``line-free'' windows used in the power-law fitting are shown with orange lines in the top panel.
Examples of the wavelength spans of the power-law-fitting bands are shown in the middle panel for a redshift of 2 and 0.5.
The bottom panel shows the $\alpha$--$z$ scatter plot of the BLAGN1 (orange) and BLAGN2 (green).
The black line ($\alpha=-0.33z$) corresponds to $\alpha'=0$.
}
}
\label{fig:eachspec}
\end{figure}

We show the BOSS spectral shape of our sources in Fig.~\ref{fig:eachspec}.
Although all the sources are defined as BLAGN, whose optical spectral shapes are expected to be a blue power-law with a negative slope around $-1.5$ \citep{VandenBerk2001}, we find that a fraction of them show continuum reddening.
To evaluate the reddening, we define a slope parameter as follows.
Since our sources span a wide redshift range, we define the slope on a redshift-dependent rest-frame band.
We choose a serial of ``line-free'' sections between rest-frame 1670 and 4800$\angstrom$
\footnote{Throughout this paper, the wavelengths correspond to rest frame if not explicitly specified.}
, excluding the line-dominated part but as little as possible, as shown in the top panel of Fig.~\ref{fig:eachspec}.
In these selected sections, for each source, we choose the bluest available part which spans a wavelength width of $0.25$ dex (see examples in the middle panel of Fig.~\ref{fig:eachspec}) to define the slope parameter.
After shifting each spectrum to rest frame, we calculate a slope $\alpha$ by a linear fitting in the selected bands.

As shown in the bottom panel of Fig.~\ref{fig:eachspec}, we find an anti-correlation between $\alpha$ and redshift.
This anti-correlation is likely caused by the R band magnitude selection bias against dust extincted sources at high redshifts.
Because dust extinction affects only the blue band in rest frame, so that the observed R band flux is less affected for low-$z$ sources than for high-$z$ sources.
To skirt around this bias, we define a less-redshift-dependent slope $\alpha'$ by applying a redshift correction $\alpha+0.33\times z$ using the slope of the anti-correlation $0.33$, as shown in the bottom panel of Fig.~\ref{fig:eachspec}.
Using this $\alpha'$ parameter, we can separate the reddened sources which appear different from the majority of the sample at different redshifts.
Hereafter, the sources with $\alpha'>0$ are called red AGN, and the others are called blue.
The top and middle panels of Fig.~\ref{fig:eachspec} clearly show the differences between their continuum shapes.

Comparing the median composite spectra (generated using method ``A'' as described in \S~\ref{sec:stackingmethod}) between the blue and red BLAGN, we get an $E(B-V)=0.27$.
It corresponds to $A_V=R_V\times E(B-V)=1.4$, if we consider the possibility that AGN might have larger dust grain size \citep[e.g.,][]{Laor1993,Maiolino2001a,Imanishi2001} and hence adopt $R_V=5.3$ \citep{Gaskell2004} as opposed to the Galactic value of 3.1 \citep[but see also][]{Weingartner2002,Willott2005}.
Note that this value only corresponds to a fraction of low-$z$ sources at $z\lesssim0.7$.
Such low-z sources have significant stellar contamination (see \S~\ref{sec:lines}), which could flatten the spectra at $\gtrsim 4000\angstrom$.
Meanwhile, as discussed above, high-z sources show lower extinction because of sample selection effects \citep[see also][]{Willott2005}.
Therefore, for the whole red AGN sample, this $A_V$ value is more of a moderate upper limit than a typical value of the optical extinction.
Adopting an empirical correlation \N{H} $= 2.2\times10^{21} A_V$ \citep{Guver2009}, it corresponds to an \N{H} of $3\times10^{21}$~cm$^{-2}$ -- approximately the lower limit of the X-ray \N{H} of the BLAGN2.
Nevertheless, red AGN only constitute a small fraction of the BLAGN2.
In most of the BLAGN2 which are blue (below the black line in the bottom panel of Fig.~\ref{fig:eachspec}), there is rarely any optical dust extinction.
Therefore, we conclude that, the dust accountable for the optical extinction is insufficient to explain the X-ray obscuration in the BLAGN2.

\section{The X-ray Obscuration}
\subsection{The Effective Eddington Limit}
\label{sec:effedd}
\begin{figure}[htbp]
\epsscale{1}
\plotone{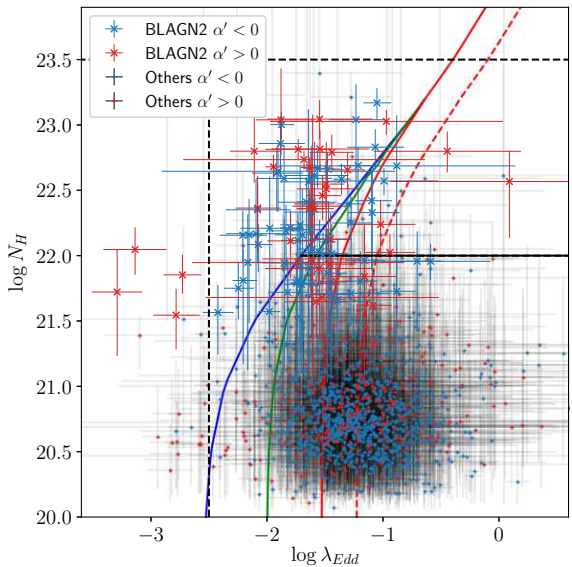}
\caption{\N{H}--\lbd{} scatter plot of the BLAGN.
The x-crosses indicate BLAGN2, and the plus markers indicate the others, including BLAGN1 and a few low S/N sources.
The sources with blue ($\alpha'<0$) and red ($\alpha'>0$) optical continua are plotted in blue and red colors, respectively.
We show the effective Eddington limit estimated by \citet{Fabian2009} with dust grain abundances of 1 (blue), 0.3 (green), and 0.1 (red).
The red dashed line shows a factor of 2 increase in the effective Eddington limit with a grain abundance of 0.1 due to the mass of intervening stars.
The black solid line corresponds to the lower boundary of $10^{22}$~cm$^{-2}$.
The black dashed lines correspond to \N{H}$=10^{23.5}$~cm$^{-2}$ and \lbd{}$=10^{-2.5}$.
}
\label{fig:NH_Edd_forbid}
\end{figure}

The effective Eddington limit is much lower for dusty gas than for ionized dust-free gas \citep{Laor1993, Scoville1995,Murray2005}.
Such a limit defines a blow-out region (forbidden region) in the \N{H}--\lbd{} plane for AGN, in which the living time of an AGN is expected to be short.
In Fig.~\ref{fig:NH_Edd_forbid}, we plot our sources in the \N{H}--\lbd{} diagram to see if they obey the effective Eddington limit.
The \lbd{} is calculated from $L_X$ assuming a constant bolometric correction factor of 20, as done in \citet{Ricci2017}.
We show the effective Eddington limits calculated by \citet{Fabian2009} for dusty gas located close to the black hole, where the black hole dominates the mass locally, with dust grain abundances of 1, 0.3, and 0.1.
As done in \citet{Ricci2017}, we plot a lower boundary of \N{H}$=10^{22}$~cm$^{-2}$, below which the obscuration might be due to galaxy-scale dust lanes.

Compared with the BLAGN1 at the same \lbd{}, a lack of BLAGN2 can be seen in the dust blow-out region at \lbd{}$\gtrsim10^{-1}$, similar as found by \citet{Ricci2017}.
Using the BAT 70-month AGN catalog, \citet{Ricci2017} found that among 160 NLAGN with \N{H}$\geqslant10^{22}$~cm$^{-2}$, a very small fraction (five sources, $3\%$) lie in the dust blow-out region (see their Fig.~3).
Among our BLAGN2 sample, 80 have \N{H}$>10^{22}$~cm$^{-2}$, and 18 out of the 80 ($\sim22\%$) lie in the dust blow-out region (blue and solid black lines in Fig.~\ref{fig:NH_Edd_forbid}). The fraction we find is larger.
Note that this is not a rigorous comparison, since the BAT survey and XXL survey have very different depths and different selection limits.
However, considering that the dust column density revealed by optical extinction is insufficient to account for the X-ray obscuration (\S~\ref{Sec:specslope}), it is likely true that BLAGN2 have a larger probability to occur in the dust blow-out region than NLAGN, in the sense that the X-ray absorbers in BLAGN2 have a lower dust fraction and thus a higher effective Eddington limit.
We choose the limit with a grain abundance of 0.1 (red solid line in Fig.~\ref{fig:NH_Edd_forbid}), and correct it by a factor of 2 to account for the mass of the stars inwards from the obscuring material, as done in \citet{Fabian2009}.
This factor corresponds to a scale of a few parsec from the nucleus in the case of our Galaxy \citep{Schodel2007}.
Using this corrected limit (red dashed line in Fig.~\ref{fig:NH_Edd_forbid}), we find that there are only three sources in the blow-out region.
Incidentally, using this limit we can efficiently select sources which are likely outflowing (see Appendix.~\ref{append:outflow}).
Therefore, we argue that the X-ray obscuration in BLAGN can be well described by such an absorber with a low dust-fraction and located at about a few parsec from the black hole.

\subsection{The major difference between BLAGN1 and BLAGN2}
\label{sec:comparison}

In this section, we compare the $L_X$, \mbh{} and \lbd{} between the BLAGN1 and BLAGN2, in order to investigate which factor is the main physical driver of the difference between them.
First, to reduce sample selection bias and compare one parameter between the BLAGN1 and BLAGN2 with the others under control, we re-select the BLAGN1 and BLAGN2 samples from sample ``0''.

To compare their \mbh{} at the same $L_X$ and $z$, we select a BLAGN1 sample and a BLAGN2 sample with the same $L_X$ and $z$ distribution.
We repeatedly select the nearest BLAGN1 to each BLAGN2 in the $\log L_X$--$z$ space.
When a BLAGN2 has no more neighbor found within a maximum distance of $0.25$, it is excluded and the procedure is restarted with the reduced BLAGN2 sample.
We find that for a subsample of $94$ BLAGN2, we could repeat the nearest-point selection for eight times; in other words, we could assign eight nearest points from the BLAGN1 to each of these BLAGN2.
As shown in Fig.~\ref{fig:reselect}, the redshift distributions are significantly different between the BLAGN1 and BLAGN2 in sample ``0'' (empty histogram), because of the X-ray sample selection bias against BLAGN2 which have relatively lower observed fluxes.
In the re-selection, the excluded BLAGN2 (empty diamonds) have relatively lower $z$ and higher $L_X$ and the excluded BLAGN1 (empty circles) have relatively higher $z$ and lower $L_X$.
As a result, the selected BLAGN1 and BLAGN2 have a highly-identical $L_X$--$z$ distribution (filled points and histograms).

To select samples with the same \mbh{}--$z$ distribution between the BLAGN1 and BLAGN2, we perform a similar sample re-selection as above in the $\log$\mbh{}--$z$ space, allowing a maximum distance of $0.25$.
In this selection, the excluded BLAGN2 have relatively lower $z$ and higher \mbh{} and the excluded BLAGN1 have relatively higher $z$ and lower \mbh{}.
Similarly, we can also select samples of BLAGN1 and BLAGN2 which have the same \lbd{}--$z$ distribution, trimming off a few BLAGN2 with relatively lower $z$ and lower \lbd{} and a fraction of BLAGN1 with relatively higher $z$ and higher \lbd{}.

There are $11$ highly obscured sources with $\log$\N{H}$>23.5$ in the sample ``0''.
We note that all except one of them (with the lowest \N{H}) are excluded in the ``same $L_X$--$z$'' selection, because they have relatively higher $L_X$ than both the BLAGN1 and the other BLAGN2 at the same redshifts. 
Such high $L_X$ of them are possibly overestimations (see Appendix \ref{append:highlyobscured}).
To be conservative, we exclude such sources and focus on the others whose \N{H} and $L_X$ are better constrained by the XMM-Newton spectra.
As shown in Fig.~\ref{fig:NH_Edd_forbid}, most of the sources at \lbd{} below $10^{-2.5}$ have a red optical continuum. At such low accretion rates, the sources likely have their optical emission dominated by the host galaxy. We also exclude them from further analysis.
Applying the two additional filters (\N{H}$<10^{23.5}$~cm$^{-2}$ and \lbd{}$>10^{-2.5}$), as shown by the black dashed line in Fig.~\ref{fig:NH_Edd_forbid}, we select a sample ``1'' from sample ``0''. All the analyses below are based on sample ``1''.

Performing the sample re-selections described above on sample ``1'', we select three pairs of samples:
\begin{description}[align=right,labelwidth=2cm]
\item [``$^=L_X$''] 92 BLAGN2 and $7\times92$ BLAGN1 with the same $L_X$--$z$ distribution.
\item [``$^=$\mbh{}''] 78 BLAGN2 and $7\times78$ BLAGN1 with the same \mbh{}--$z$ distribution.
\item [``$^=$\lbd{}''] 78 BLAGN2 and $7\times78$ BLAGN1 with the same \lbd{}--$z$ distribution.
\end{description}
We compare the \mbh{} and \lbd{} between the ``$^=L_X$'' BLAGN1 and BLAGN2 samples, compare the \lbd{} and $L_X$ between the ``$^=$\mbh{}'' samples, and compare the \mbh{} and $L_X$ between the ``$^=$\lbd{}'' samples.
For each comparison we perform a K-S test \footnote{Throughout this paper, the K-S test probability denotes 1 minus the probability that two samples are drawn from the same population.}.

\begin{figure}[htbp]
\epsscale{1}
\plotone{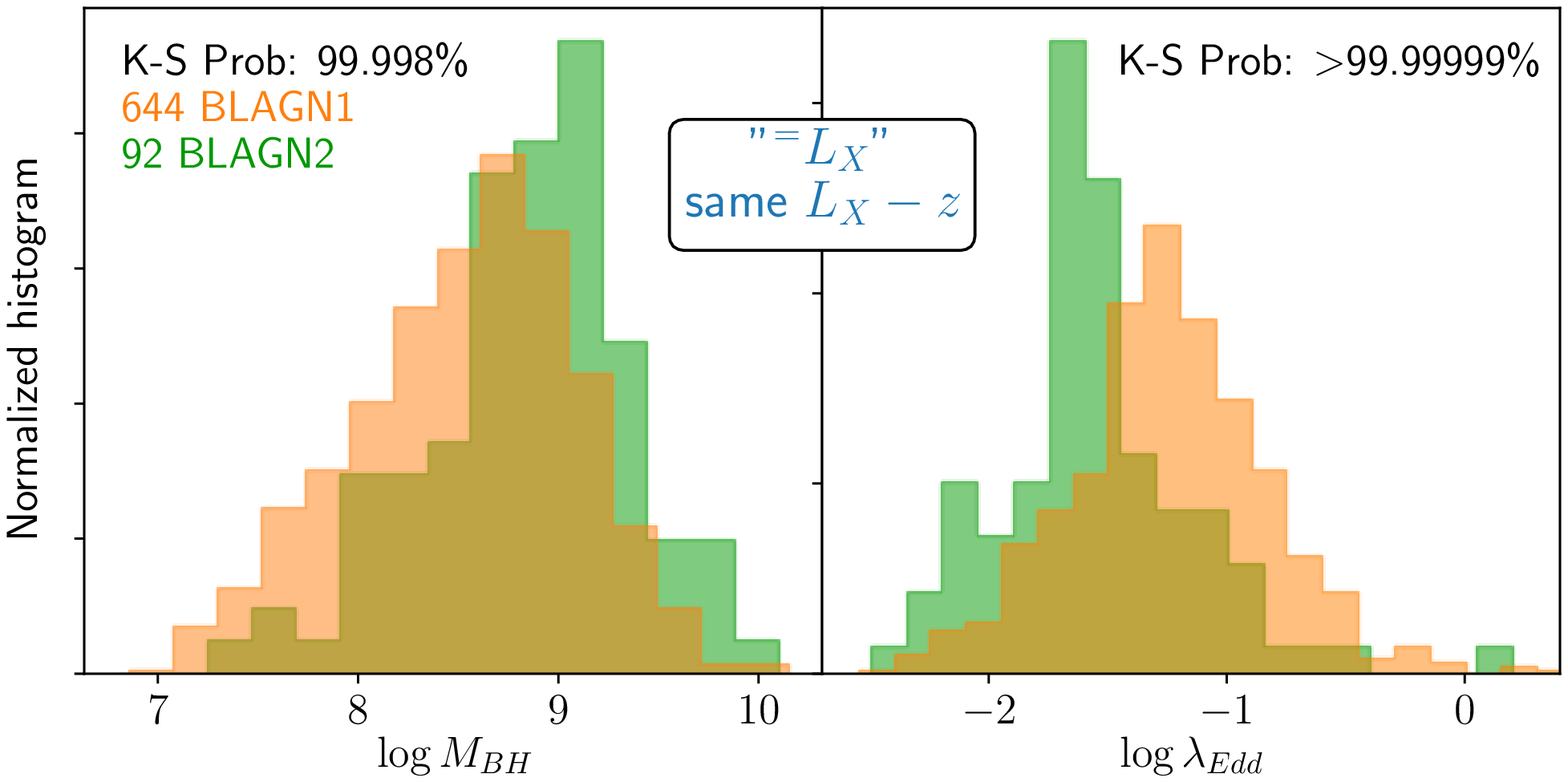}
\plotone{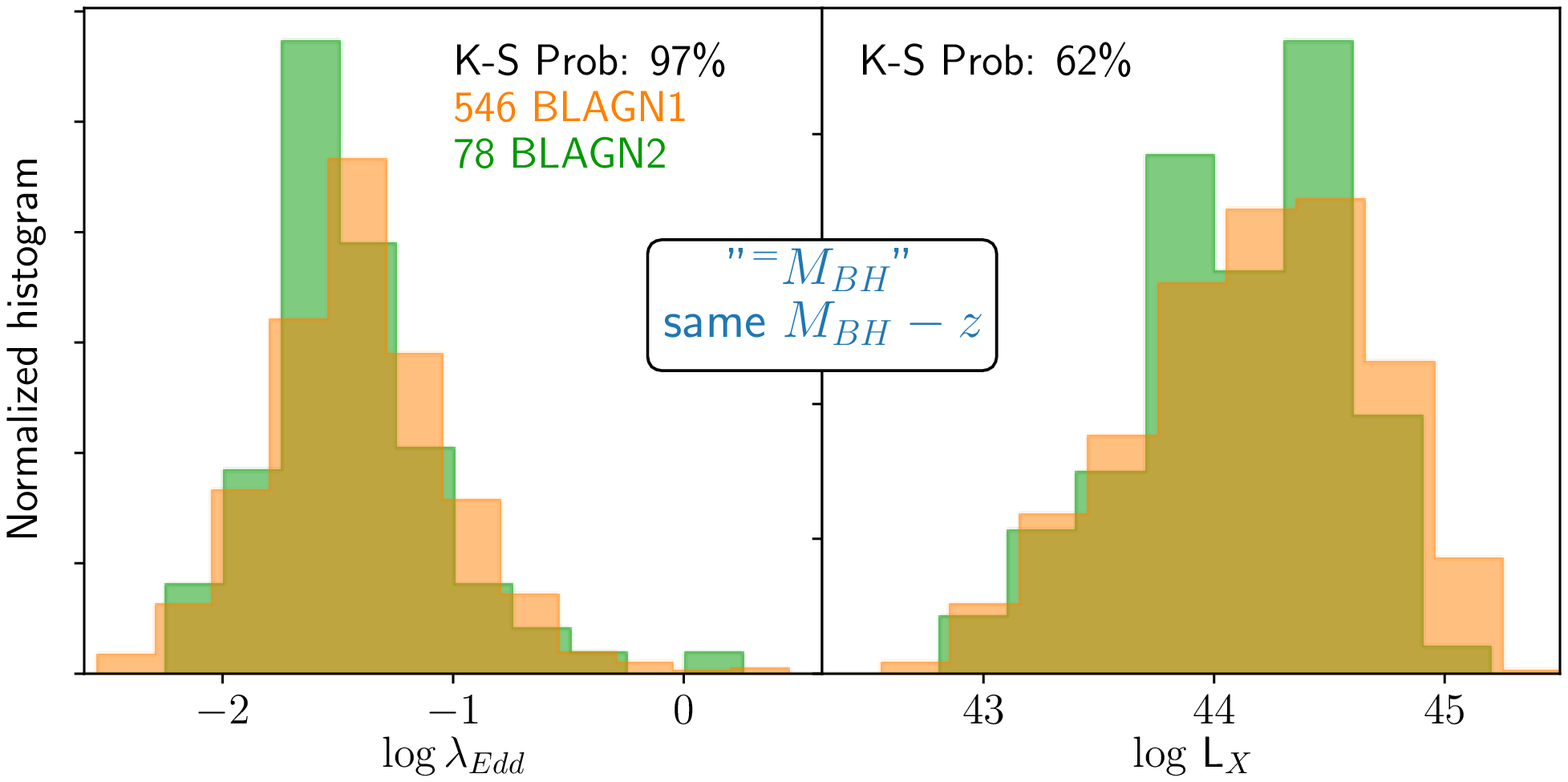}
\plotone{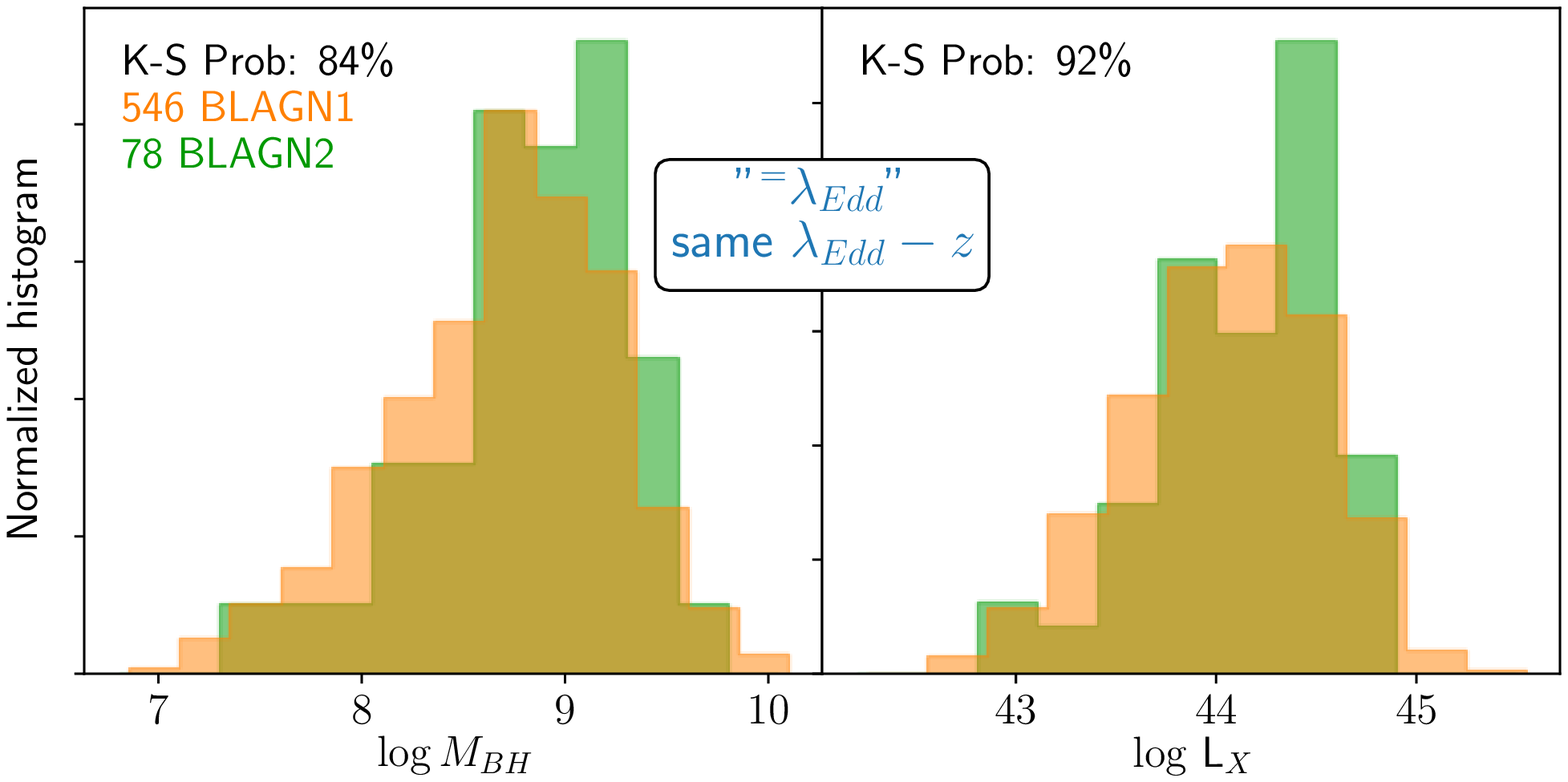}
\caption{Comparison of distributions of \mbh{}, $L_X$, or \lbd{} between the BLAGN1 and BLAGN2 using a series of samples -- ``$^=L_X$'', ``$^=$\mbh{}'', and ``$^=$\lbd{}'', from top to bottom.
Number of sources and K-S test probability are shown in each panel.
}
\label{fig:hist_comp}
\end{figure}

As shown in Fig.~\ref{fig:hist_comp}, at the same $L_X$ (top panel, ``$^=L_X$''), the BLAGN2 have significantly higher \mbh{} and lower \lbd{}.
The median \mbh{} and \lbd{} of the ``$^=L_X$'' samples are 8.60, -1.26 for the BLAGN1 and 8.93, -1.60 for the BLAGN2.
At the same \mbh{} (middle panel, ``$^=$\mbh{}''), the $L_X$ is similar between the BLAGN1 and BLAGN2; the \lbd{} is lower in the BLAGN2 but only slightly (about 2$\sigma$).
At the same \lbd{} (bottom panel, ``$^=$\lbd{}''), we find no significant differences either on the other two parameters between the BLAGN1 and BLAGN2.
Therefore, among $L_X$, \mbh{}, and \lbd{}, the main physical difference between the BLAGN1 and BLAGN2 consists in \mbh{} or \lbd{}.
However, as will be discussed in \S~\ref{sec:physicaldriver}, the \mbh{} difference can be alternatively attributed to a bias caused by an inclination effect.

A constant bolometric correction factor is used to calculate \lbd{} from $L_X$ (\S~\ref{sec:effedd}). However, if we consider the correlation between the X-ray bolometric correction factor and \lbd{} \citep{Vasudevan2007, Vasudevan2009, Lusso2012, Liu2016}, the \lbd{} of BLAGN2 can be even lower and thus the difference between the \lbd{} of the BLAGN1 and BLAGN2 can be even stronger.

\section{The Optical Spectra}
\label{Sec:OpticalSpec}

\subsection{The Optical Extinction}
\label{sec:compslopes}
\begin{figure}[htbp]
\epsscale{1}
\plotone{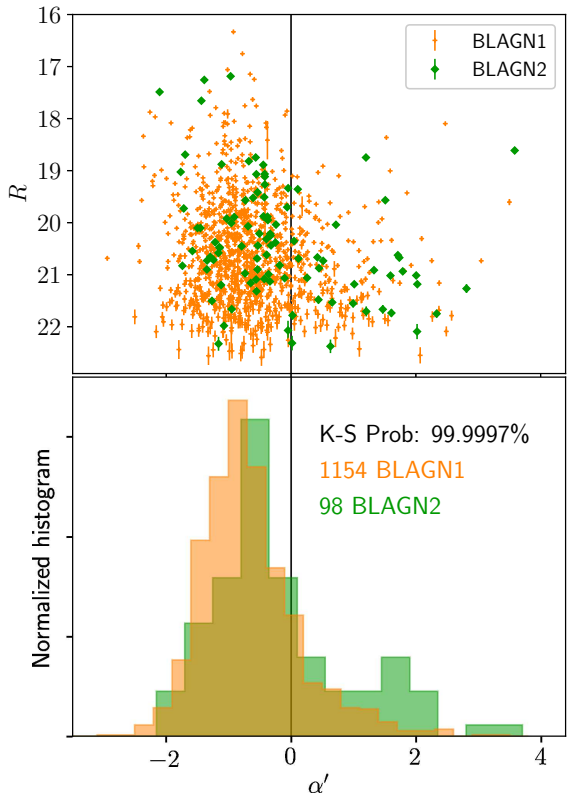}
\caption{
The scatter plot of R band magnitude $R$ and $\alpha'$ (upper panel) and the normalized $\alpha'$ distributions (lower panel) of the BLAGN1 (orange) and BLAGN2 (green) of sample ``1''.
The vertical black line corresponds to $\alpha'=0$.
The errorbars of $R$ correspond to $1\sigma$.
}
\label{fig:slopes_P}	
\end{figure}

In \S~\ref{Sec:specslope}, we have defined a slope parameter $\alpha'$ to evaluate the optical continuum reddening.
In the upper panel of Fig.~\ref{fig:slopes_P}, we plot $\alpha'$ against the observed R band magnitude for sample ``1''.
Clearly, the red ($\alpha'>0$) sources have relatively lower optical fluxes, indicating that the reddening should be mainly caused by dust extinction, which reduces the optical flux.

As a first, rough comparison between the optical continuum of the BLAGN1 and BLAGN2, we compare their slopes $\alpha'$ using the sample ``1'' in the lower panel of Fig.~\ref{fig:slopes_P}.
We find that the slopes are significantly redder in the BLAGN2 sample at a K-S test confidence probability of 99.9997\%.
Using the ``$^=L_X$'', ``$^=$\mbh{}'', or ``$^=$\lbd{}'' samples, we also find such differences at high K-S test probabilities of 99.9992\%, 99.8\%, and 99.94\%, respectively.
Therefore, the optical dust extinction is correlated with the X-ray obscuration, and this correlation is not driven by the \lbd{} or \mbh{} difference between the BLAGN1 and BLAGN2 as we find in \S~\ref{sec:comparison}.

With stronger dust extinction, the BLAGN2 can be more easily missed by the R band magnitude-limited sample selection threshold.
Taking account of this bias will boost the difference of dust extinction levels between the BLAGN1 and BLAGN2.

\subsection{Spectral Stacking Methods}
\label{sec:stackingmethod}
In order to compare the optical spectra of the BLAGN1 and BLAGN2 in more details, we stack their SDSS spectra.
Three different methods of stacking, A, B, and C, will be used below for different purposes.
In method ``A'', we calculate the geometric mean or median of the normalized spectrum of each source in order to study the continuum shape.
First, for each spectrum, we apply galactic extinction correction using the extinction function of \citet{CCM1989} and exclude the
 bins with high sky background or with observed wavelength below 3700$\angstrom$ or above 9500$\angstrom$.
Then, we shift the spectrum to rest-frame using a bin size of $1e-4$ dex, which is the same as the bin size of the original spectrum,
and select the 100 available bins with the longest wavelength below rest-frame $5100\angstrom$ as the normalizing window for each spectrum.
All the spectra are then ordered by redshift, and, starting from the second one, each spectrum is normalized in the selected window to the composite spectrum of the sources with lower redshifts.
To calculate the normalization factor, we use the best-fit models from the SDSS pipeline instead of the real spectra to avoid the high variance of the spectra in some low S/N cases (see Fig.~\ref{fig:eachspec} for an illustration of the normalization).
Using the extinction corrected, shifted, and normalized spectra, a composite spectrum (geometric mean or median) is calculated.
Then we repeat the normalizing procedure but normalizing each spectrum to the generated composite spectrum instead of normalizing each spectrum to the ones with lower redshifts.
The composite spectrum converges after a few iterations, then the $68\%$ confidence intervals are measured using the bootstrap percentile method.
Using this method, we present the composite spectrum of our BLAGN in Appendix.~\ref{append:spec}.

Stacking method ``B'' is used to compare the optical fluxes between samples.
In this method, we calculate the median of all the sources in each wavelength bin without normalizing the spectra to each other.
Instead, each extinction corrected and shifted spectrum is multiplied by $D_L^2(1+z)$, where $D_L$ is the luminosity distance, in order to preserve the luminosity.
The $16\%$ and $84\%$ percentile spectra are used to estimate the flux scatter.
Note that such a composite spectrum does not represent the spectral shapes of the sources, it shows the fluxes of the sources at specific redshifts -- low-$z$ sources dominate the red part and high-$z$ sources dominate the blue part.

To measure the line features in the composite spectra, we use a stacking method ``C'', which calculates the median spectra of the ratios of each spectrum to its local best-fit continuum.
For two sets of lines -- \Ne{III}--\Ca{II} and H$\beta$--\Oline{III} -- we fit the local continuum of each source in two different sets of bands -- $3750\sim3800$,$3885\sim3910$,$3947\sim3960$, and $3990\sim4050$ for the former, $4600\sim4750$ and $5050\sim5200$ for the latter.
Rather than measuring line EW accurately, we aim at making a comparison of the line EW between different types of sources.
Therefore, we just fit the spectra with a simple power-law in the selected windows and calculate the data to model ratio.
The median of the ratios are calculated as the composite line spectrum.

When studying the continuum with method ``A'' and ``B'' as above, we bin the spectra by a factor of $8$ to reduce fluctuation.
To study the line features, we use the unbinned spectra with the original bin size of $1e-4$ dex.
Sources at different redshifts contribute to the composite line spectra for different line sets .
If a source has less than 10 available bins on the left or right side of the \Ca{II} or \Oline{III} line, it is excluded from the stacking of the corresponding line set.

\subsection{The Optical Continuum}
\label{sec:continuum}

\begin{figure}[htbp]
\begin{center}
\epsscale{1}
\plotone{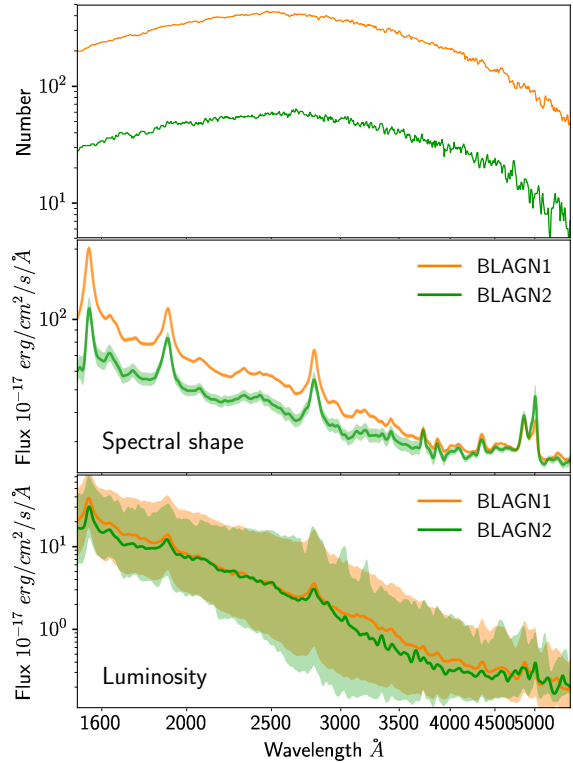}
\caption{
The top panel shows the number of contributing sources per bin.
The middle panel shows the geometric mean spectra and $1\sigma$ error of the BLAGN1 (orange) and BLAGN2 (green) generated using method ``A'', which represent the mean spectral shapes.
The bottom panel shows the composite spectra generated using method ``B'' and the $68\%$ scatter of the samples, which represent the mean luminosities.
The ``$^=$\mbh{}'' samples are used.
}
\label{fig:spec}
\end{center}
\end{figure}

To compare the optical continuum shape between the BLAGN1 and BLAGN2, we stack the ``$^=$\mbh{}'' BLAGN1 and BLAGN2 spectra respectively using method ``A''.
As shown in Fig.~\ref{fig:spec}, having the same redshift distribution, the curves of source number per bin have the same shape for the BLAGN1 and BLAGN2 samples (the top panel). 
The composite spectrum of BLAGN2 is much flatter than that of BLAGN1 (the middle panel), showing the higher probability of continuum reddening in BLAGN2.

In order to check whether the optical luminosities of BLAGN2 are reduced by dust extinction, we compare the composite spectra generated using method ``B'' between the ``$^=$\mbh{}'' BLAGN1 and BLAGN2 samples. As shown in the bottom panel of Fig.~\ref{fig:spec}, we find no significant difference. It is because the dust extinction occurs in both BLAGN1 and BLAGN2 and in only a small fraction of them. Taking also the large luminosity scatter into account, the dust extinction could not reduce the mean luminosity of the whole BLAGN1 or BLAGN2 sample significantly.

Comparing the composite spectra between the BLAGN1 and BLAGN2 using the ``$^=L_X$'' or ``$^=$\lbd{}'' samples, we find similar results about both the spectral shape and the optical luminosity.
As discussed in \S~\ref{Sec:OpticalSpec}, the more severe continuum reddening in BLAGN2 than in BLAGN1 is not driven by differences in \lbd{} or \mbh{}; it is just associated with the X-ray obscuration.

\subsection{The Line Features}
\label{sec:lines}

\begin{figure*}[htbp]
\epsscale{1.5}
\plotone{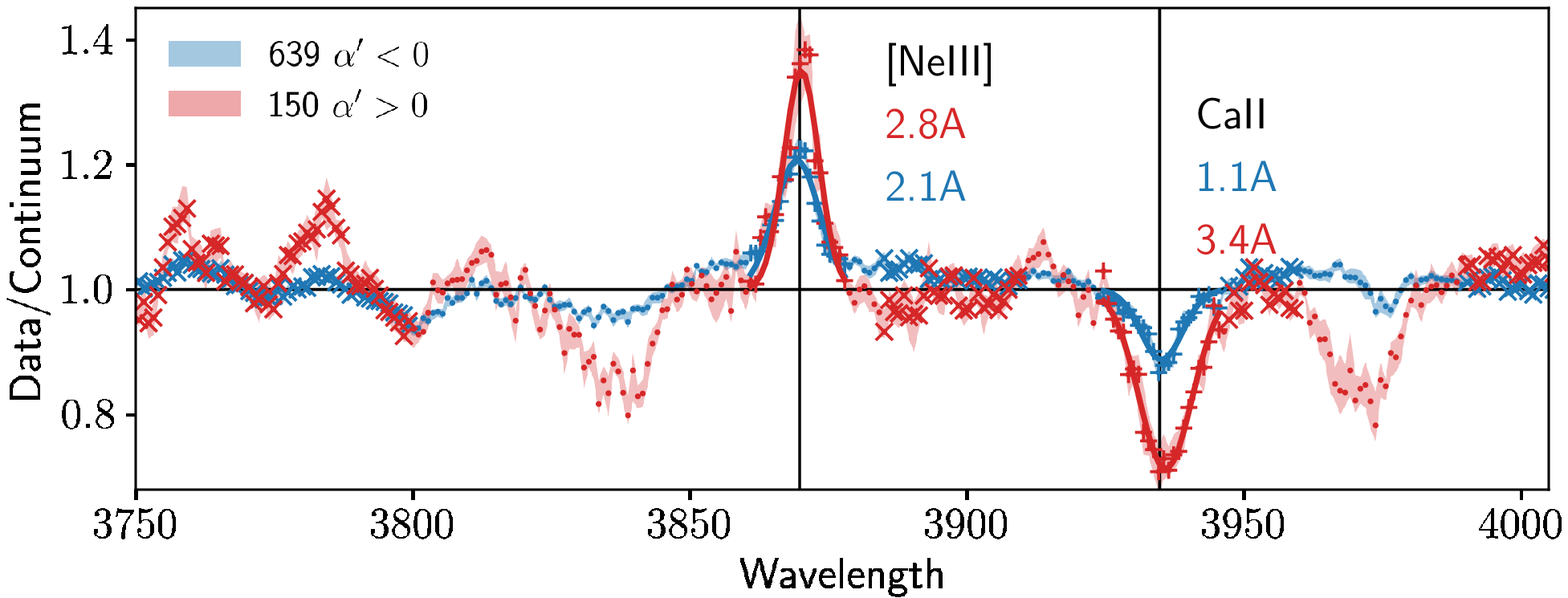}
\plotone{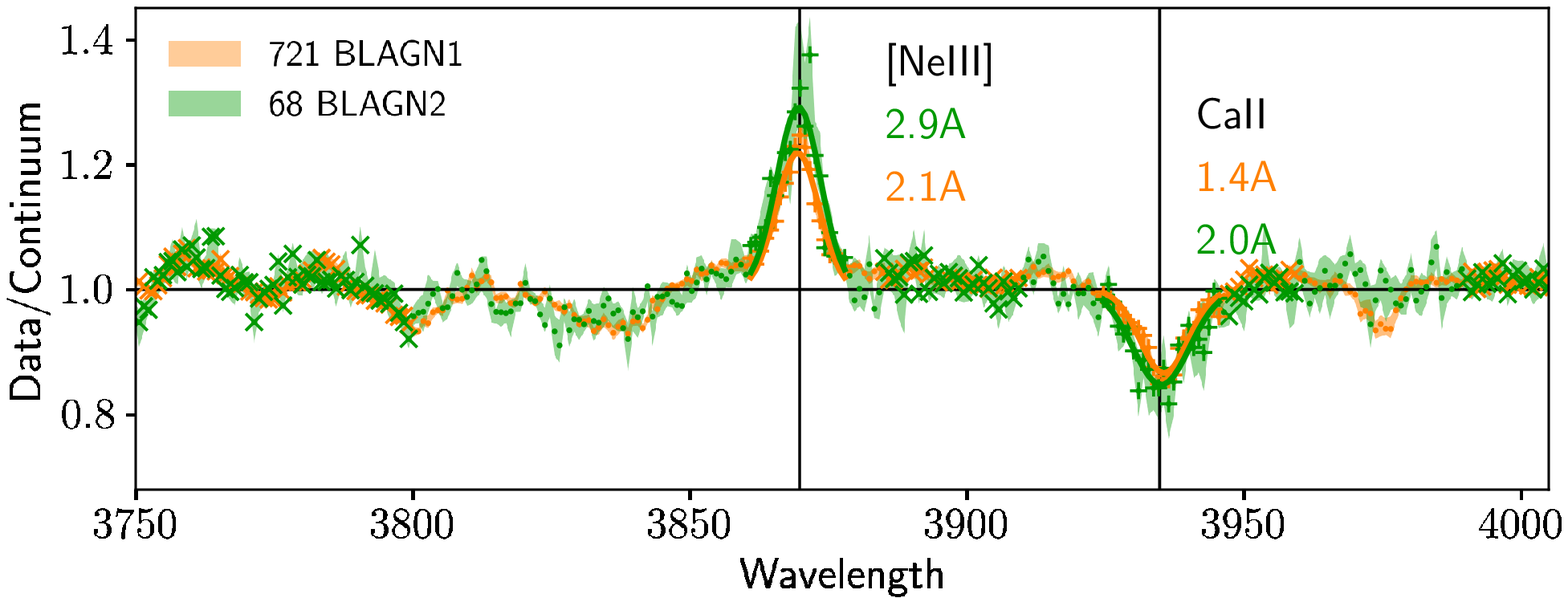}
\plotone{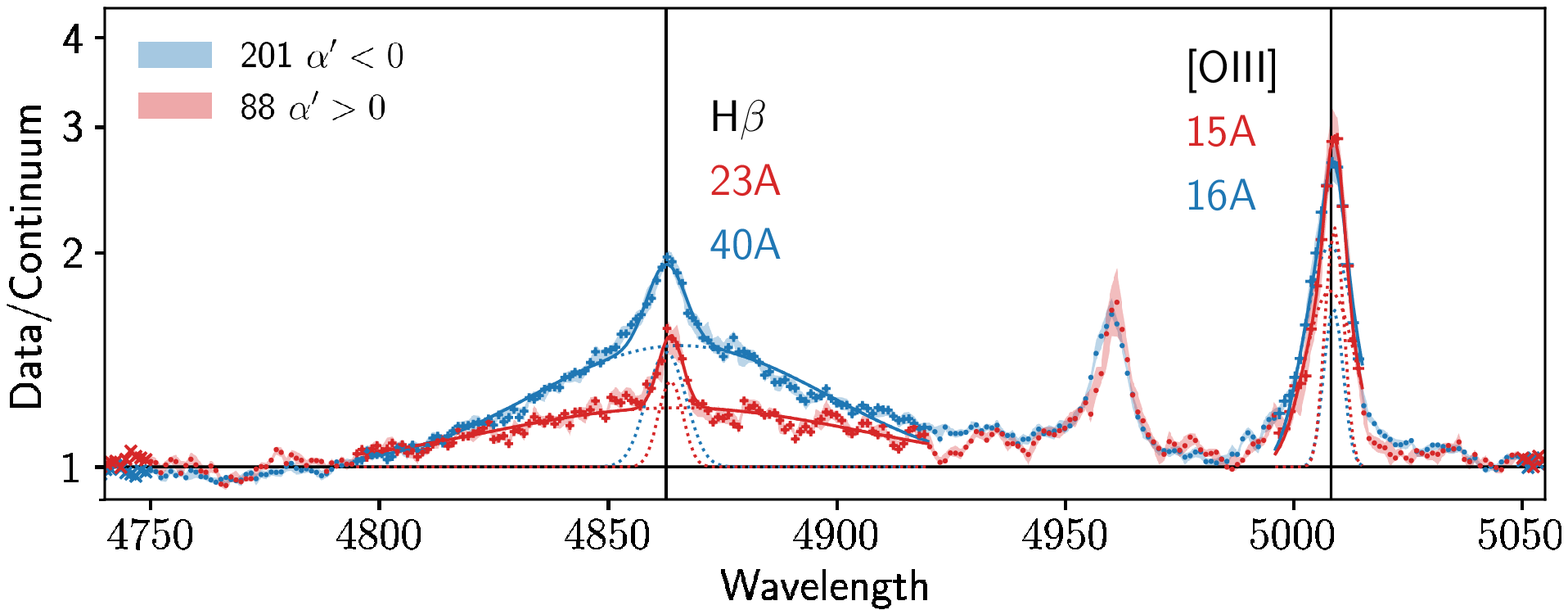}
\plotone{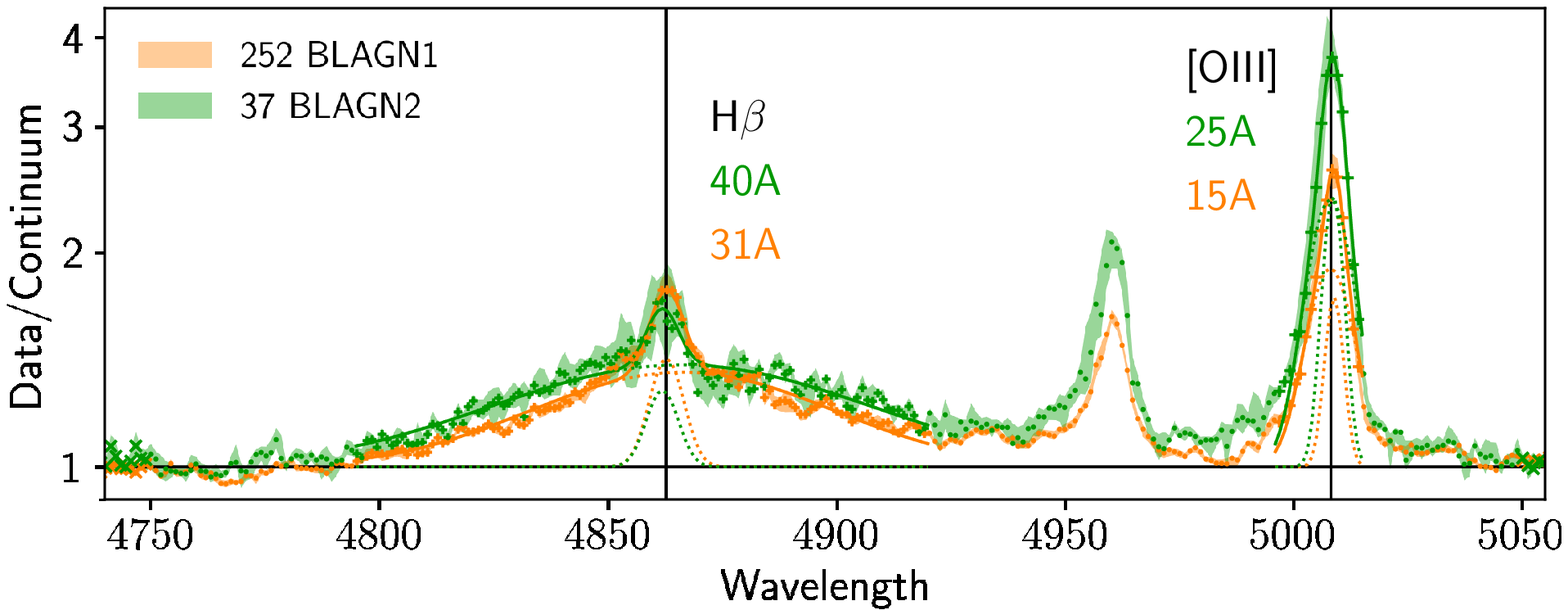}
\caption{
Comparison of the composite line spectra between the blue ($\alpha'<0$, blue color) and red ($\alpha'>0$, red color) AGN and between the BLAGN1 (orange) and BLAGN2 (green), using sample ``1''.
The upper two panels show the \Ne{III} 3869 emission line and the \Ca{II} K 3934 absorption line.
The lower two panels show the broad H$\beta$ and narrow \Oline{III} lines.
The EW of the lines are marked in the figure using corresponding colors.
The best-fit profiles (single- or double-gaussian) are plotted with solid lines.
In the cases of double-gaussian models (for H$\beta$ and \Oline{III}), each individual gaussian profile is plotted with dotted lines.
The data points involved in the line fitting are marked with plus crosses.
The points involved in the local continuum fitting are marked with x-crosses.
The vertical lines correspond to the rest-frame wavelengths the lines.
}
\label{fig:alllines}
\end{figure*}

We have seen in the previous sections that the X-ray obscuration in BLAGN2 is statistically associated to a higher level of optical spectral reddening, which is likely caused by dust extinction.
In order to further understand the reason of the reddening, we compare the line features, not only between the BLAGN1 and BLAGN2, but also between the red ($\alpha'>0$) and blue ($\alpha'<0$) BLAGN.
This is because the BLAGN1 and BLAGN2 samples have highly overlapped dust reddening distributions (Fig.~\ref{fig:slopes_P}), but the red and blue BLAGN samples separate the sources with relatively low and high reddening levels distinctly.
Note that, at low redshifts, the red BLAGN sample also tend to select sources with strong stellar contaminations in the optical spectra.
In this section, we divide the sample ``1'' into red and blue subsamples and into BLAGN1 and BLAGN2 samples, and then make the composite line spectra for two sets of lines -- \Ne{III}--\Ca{II} and H$\beta$--\Oline{III} -- using method ``C'' (see \S~\ref{sec:stackingmethod}) for each of the four samples.
Using the ``$^=L_X$'', ``$^=$\mbh{}'', or ``$^=$\lbd{}'' samples instead does not change the results obtained in this section, because the optical spectral difference between the BLAGN1 and BLAGN2 is not driven by any physical parameters ($L_X$, \mbh{}, or \lbd{}), as noticed previously in \S~\ref{sec:compslopes} and \ref{sec:continuum}.
The composite line spectra are shown in Fig.~\ref{fig:alllines}.

To estimate the line EW, we fit the narrow \Ne{III} and \Ca{II} lines with single-gaussian profiles and fit the H$\beta$ and \Oline{III} 5007 lines with double-gaussian profiles (see Fig.~\ref{fig:alllines}).
The line EW can be affected by two factors in opposites ways: dust extinction enhances line EW by reducing the underlying continuum and stellar contamination reduces line EW by enhancing the underlying continuum.
For each pair of lines, the two lines have similar wavelengths, so that the impact of dust extinction should be similar.
Suppose the local continuum of the blue BLAGN (BLAGN1) is composed of a power-law component $p$ and a galaxy component $g$, and in the red BLAGN (BLAGN2), the power-law emission is reduced by a factor of $1-a$ ($a<1$), the galaxy emission is increased by a factor of $1+b$, and the narrow line flux remains the same.
With respect to the blue BLAGN (BLAGN1), the EW of AGN emission line in the red BLAGN (BLAGN2) is enhanced by a factor of $\frac{p+g}{p-ap+g+bg}$, and the EW of stellar absorption line is enhanced by a factor of $\frac{(p+g)(1+b)}{p-ap+g+bg}$.

The first pair of lines to consider are the AGN emission line \Ne{III} 3869 and the galaxy absorption line \Ca{II} K 3934 (the upper two panels of Fig.~\ref{fig:alllines}).
We find that, compared with the blue BLAGN (BLAGN1), both lines are enhanced in the red BLAGN (BLAGN2).
It indicates that the impact of dust extinction is stronger than that of stellar contamination ($ap>bg$).
In the comparison between the blue and red BLAGN, the enhancement amplitude of the stellar absorption line (\Ca{II}) is much larger than that of the AGN emission line (\Ne{III}), indicating that the stellar contamination is strong ($1+b$ is significantly $>1$).
It is not the case in the comparison between the BLAGN1 and BLAGN2, indicating that the stellar components are similar between them ($b\ll1$).

The second pair of lines to consider are the broad H$\beta$ line and the narrow \Oline{III} 5007 line.
We find that they are enhanced in the BLAGN2 with respect to the BLAGN1, but not in the red BLAGN with respect to the blue BLAGN.
It is because at such long wavelengths ($\sim 5000\angstrom$), the relative strength of the stellar component ($g/p$) is much larger than at $\sim 4000\angstrom$, so that in the latter case (blue vs. red BLAGN) stellar contamination effect becomes strong enough to counteract the dust extinction effect ($bg\gtrsim ap$).
In the former case, the stellar contamination does not make a significant difference between the BLAGN1 and BLAGN2 ($b\ll1$) and the major difference consists in the dust extinction.
The impact of extinction at the \Oline{III} wavelength can be stronger than at $\sim 4000\angstrom$, because the shorter-wavelength section corresponds to sources at higher redshifts, where the sample selection biases against sources with high extinction levels (see the bottom panel of Fig.~\ref{fig:eachspec}).

In the meanwhile, we find that the relative strength (EW ratio) of the H$\beta$ broad line with respect to the \Oline{III} narrow lines is weaker in the red BLAGN than in the blue ones and also weaker in the BLAGN2 than in the BLAGN1. In other words, at the same \Oline{III} luminosity, the broad H$\beta$ line luminosity is lower when dust extinction occurs (see Appendix.~\ref{append:outflow} for examples). It indicates that the optical absorber of the accretion disc could partially block the BLR.

\begin{figure}[htbp]
\epsscale{1}
\plotone{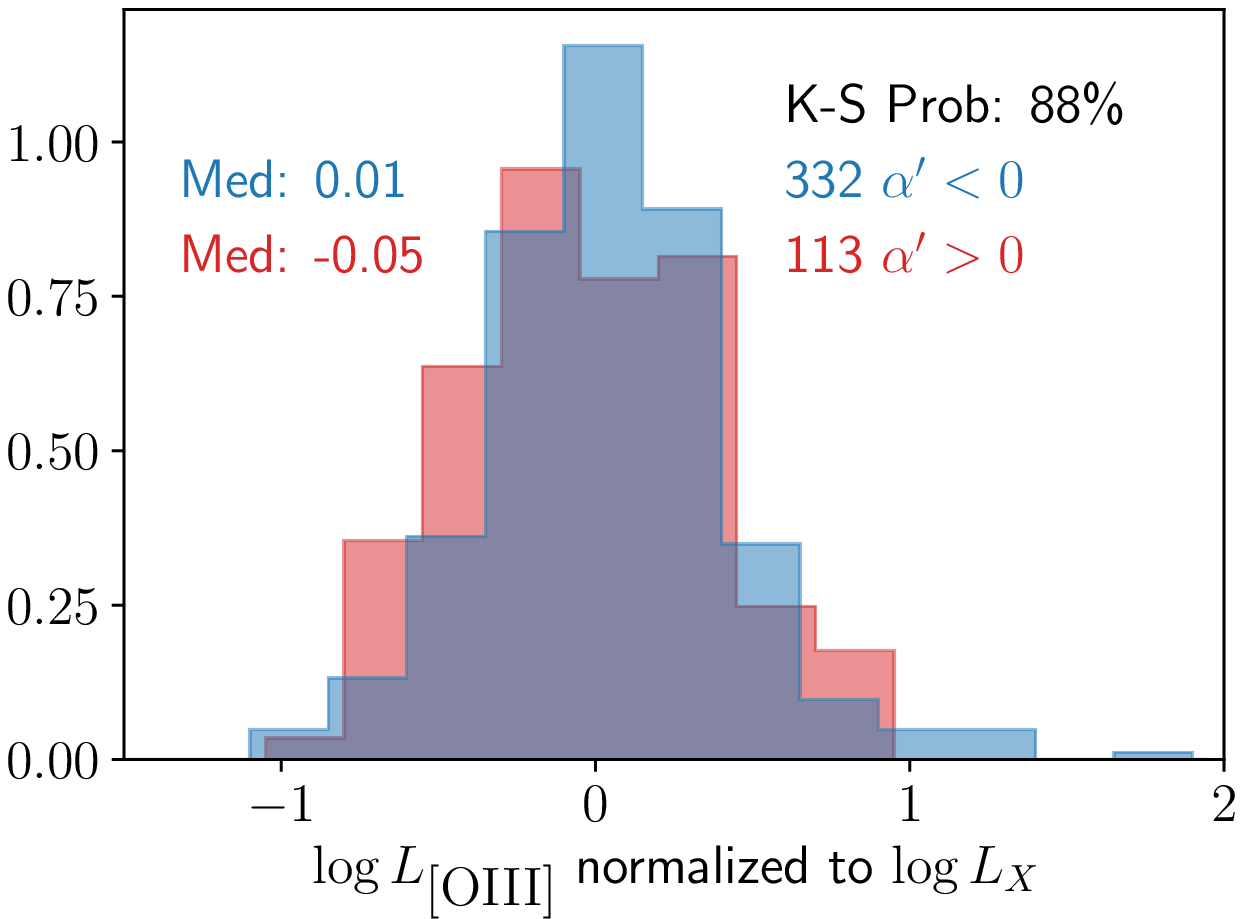}
\plotone{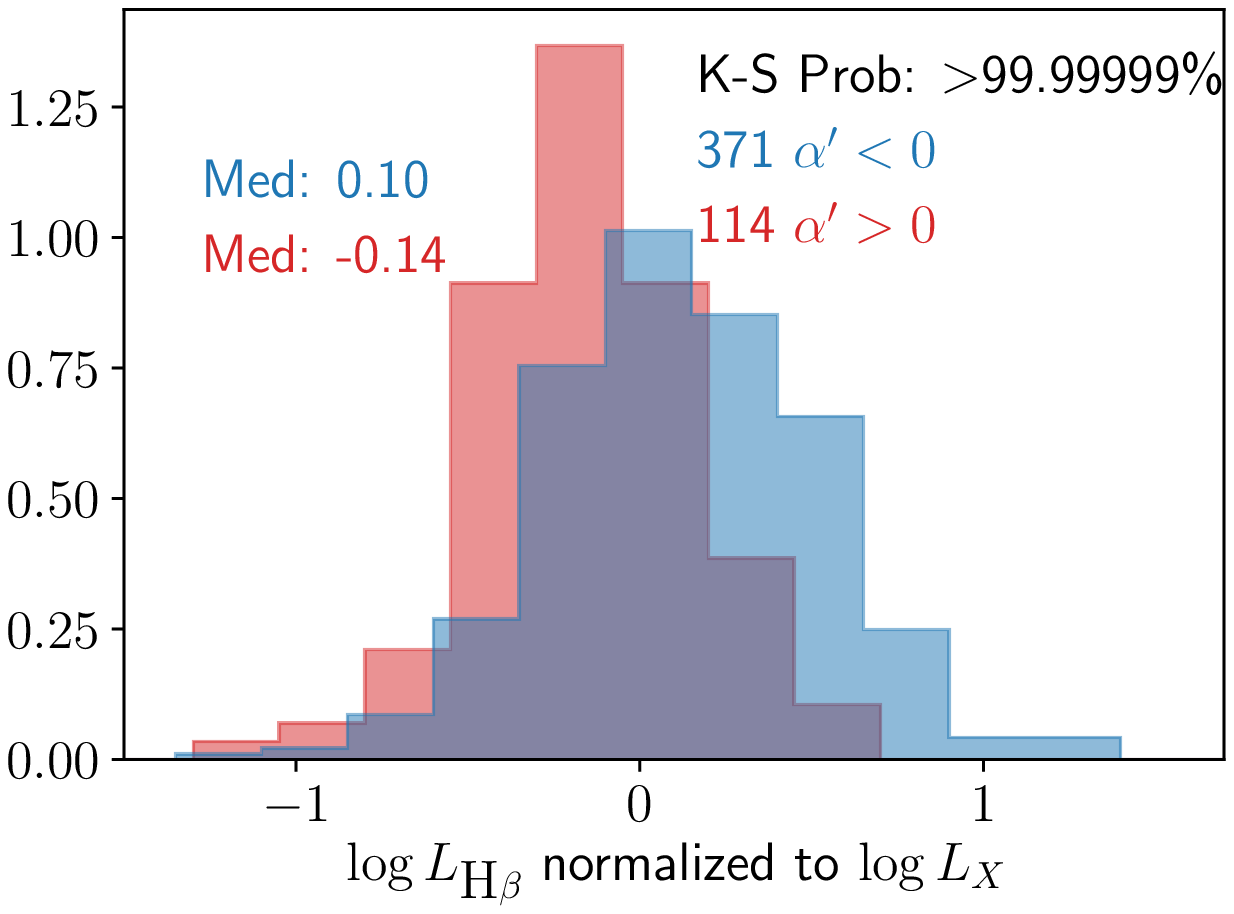}
\plotone{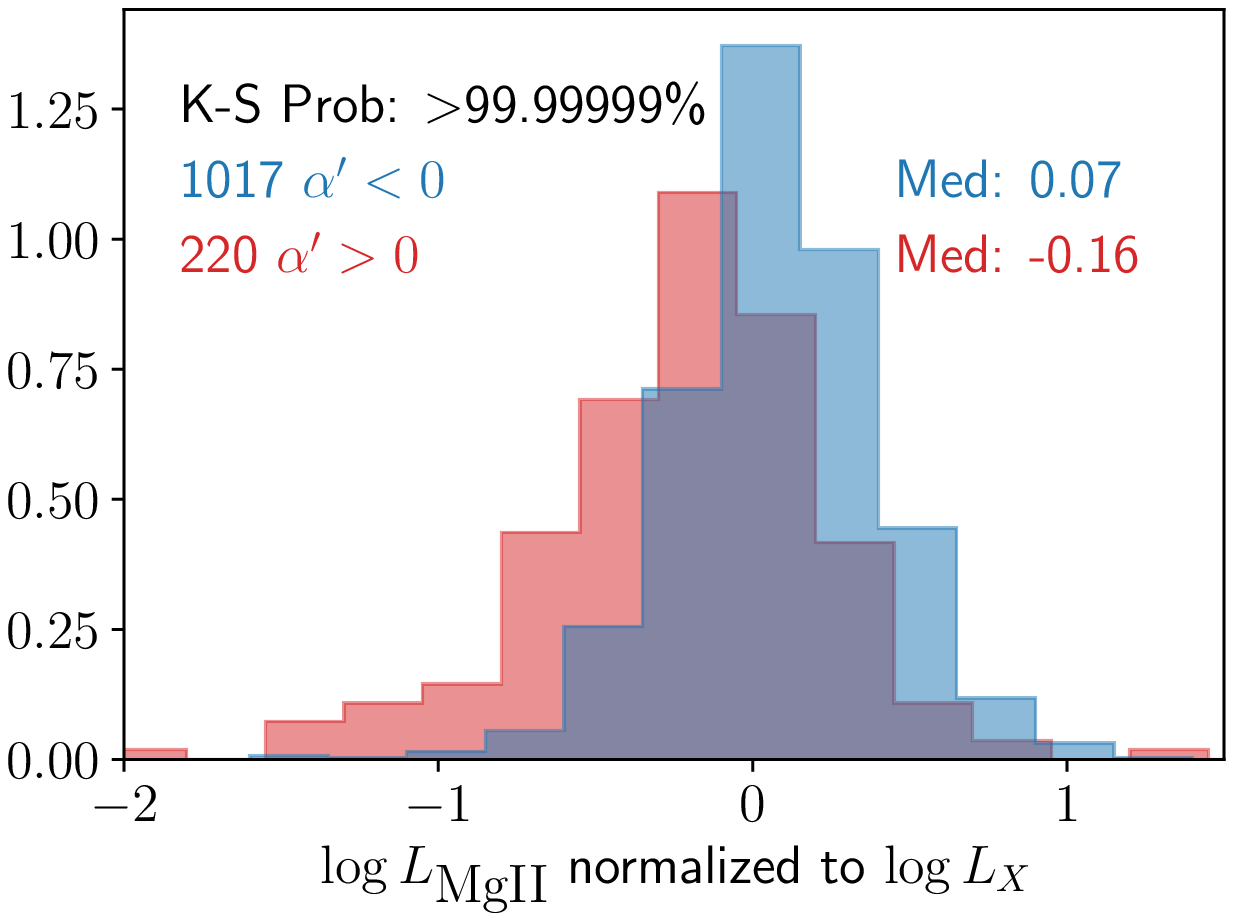}
\caption{
Comparisons of the line luminosities normalized to $L_X$ for \Oline{III}, H$\beta$, and \Mg{II} between the blue ($\alpha'<0$) and red ($\alpha'>0$) BLAGN using sample ``1''.
We mark the sample sizes, median values, and K-S test probabilities in the figure.
}
\label{fig:line2X}
\end{figure}

To test the possibility of partially blocked BLR, we calculate the relative strength of the \Oline{III}, H$\beta$, and \Mg{II} lines with respect to X-ray as the deviation of the $\log L_{line}$ from the best-fit line of $\log L_{line}$--$\log L_X$, using the $L_{line}$ from the DR9 quasar catalog built by \citet{Shen2011} as an extension of the DR7 catalog.
The relative line strength are compared between the blue and red BLAGN in Fig.~\ref{fig:line2X}.
We find that, for the narrow \Oline{III} line, the relative strength is similar between the blue and red AGN.
For the broad H$\beta$ and \Mg{II} lines, the relative strength is significantly weaker in the red AGN than in the blue ones.
A similar comparison between the BLAGN1 and BLAGN2 do not show any significant difference because of a few reasons -- the highly overlapped extinction levels of the BLAGN1 and BLAGN2 (Fig.~\ref{fig:slopes_P}), the small sample size of the BLAGN2, and the extra scatter introduced by the $L_X$.

We conclude that, in some BLAGN, the optical absorber could partially block the BLR. 
In this sense, the dust extinction in BLAGN is similar to the case of NLAGN, where an absorber at a scale between the BLR and the NLR blocks the former and not the latter.
As shown in Fig.~\ref{fig:alllines}, the H$\beta$ EW is larger in BLAGN2 than in BLAGN1, because the higher dust extinction in BLAGN2 reduces the continuum more significantly than the broad line.

The \mbh{} of the BLAGN is calculated on the basis of the broad line FWHM and optical continuum luminosity.
Practically, the continuum luminosity is substituted with broad line luminosity \citep{Shen2011}.
Therefore, the partial-covering of BLR in the BLAGN2 could cause an underestimation of their \mbh{}.
This bias can not be strong, considering that the difference of the relative broad line strength between the BLAGN1 and BLAGN2 can only be revealed in terms of the ratio of the median H$\beta$ EW to the median \Oline{III} EW (Fig.~\ref{fig:alllines}) but not in terms of the relative broad line luminosity to X-ray luminosity.
However, taking this into account will boost the difference we find between the $M_{BH}$ or \lbd{} of the BLAGN1 and BLAGN2.

\section{Conclusion and Discussion}
On the basis of the XMM-Newton X-ray spectra analysis in the XMM-XXL survey and the optical spectroscopic follow-up of the XXL sources in the SDSS-BOSS survey, we compare the BLAGN1 and BLAGN2 to study their X-ray obscuration and related properties. The results are summarized and explained as follows.

\subsection{The X-ray Absorber}
We find that, at the same $L_X$, BLAGN2 have significantly higher \mbh{} and lower \lbd{} than BLAGN1; while at the same \mbh{} or \lbd{}, no significant difference about $L_X$ is found between BLAGN1 and BLAGN2 (Fig.~\ref{fig:hist_comp}).
In other words, the major difference between BLAGN1 and BLAGN2 consists in \mbh{} or \lbd{} and not in $L_X$.
In the space of \N{H}--\lbd{}, we find a significant lack of BLAGN2 above the effective Eddington limit of a low dust fraction, where the absorber can be swept out by radiation pressure.
These properties of the X-ray absorbers in BLAGN are similar as those in NLAGN \citep{Ricci2017}.

One possibility to explain the X-ray obscuration in BLAGN is to cut off the relation between the non-simultaneous X-ray and optical observations by invoking a small X-ray obscuring cloud, which has moved away during the optical follow-up or being too small to block the extended optical emitting region (disc and BLR) ever.
However, the significant difference of \mbh{} and \lbd{} between the BLAGN1 and BLAGN2 indicates that such a possibility could only be a minor factor and there should be an intrinsic difference between them.

Unlike the optically-thick dust component in NLAGN, whose column density is too high to be measured by means of transmitted optical emission, the optical dust extinction in BLAGN is thin and occasional.
Such a thin dusty absorber is far from enough to explain the X-ray absorption in the BLAGN2 (\S~\ref{Sec:specslope}).
The line-of-sight absorbers in the BLAGN2 must have a low overall dust fraction (\S~\ref{sec:effedd}), either in terms of a low dust-to-gas ratio, or in terms of a multi-layer absorber composed of an inner gas component and an outer dust component.
Meanwhile, as revealed by IR emission, the dust column densities in NLAGN also appear lower than the X-ray obscuring column density \citep[e.g.,][]{AlonsoHerrero1997,Granato1997,Fadda1998,Georgantopoulos2011,Burtscher2016}.
Therefore, in both NLAGN and BLAGN2, the X-ray absorption is at least partially due to a dust-free component.

\subsection{The Optical Absorber}

A small fraction of the BLAGN show optical continuum reddening caused by dust extinction.
The reddening occurs in both BLAGN1 and BLAGN2, however, BLAGN2 have a higher probability to be reddened than BLAGN1 (Fig.~\ref{fig:slopes_P}), giving rise to a flatter composite spectrum of BLAGN2 than that of BLAGN1 (Fig.~\ref{fig:spec}).

The median EW of a few optical line features, as measured through composite spectra, are compared between the optical red ($\alpha'>0$) and blue ($\alpha'<0$) BLAGN and between the BLAGN1 and BLAGN2 (Fig.~\ref{fig:alllines}).
We find that, in the case of red versus blue BLAGN, both dust extinction and stellar contamination affect the line EW.
In the case of BLAGN1 versus BLAGN2, stellar contamination does not make a significant difference between them and the major difference consists in the higher dust extinction level in the BLAGN2, which enhances the line EW in BLAGN2 with respect to BLAGN1.

Using the median EW of the broad H$\beta$ line and the narrow \Oline{III} line, we find that the relative strength of H$\beta$ with respect to \Oline{III} is weaker in the red AGN (BLAGN2) than in the blue AGN (BLAGN1)  (Fig.~\ref{fig:alllines}).
We also find that the relative strength of the broad H$\beta$ and \Mg{II} line luminosities with respect to the X-ray luminosities are weaker in the red AGNs than in the blue AGNs (Fig.~\ref{fig:line2X}).
They indicate a partial-covering obscuration on the BLR.

To summarize, the X-ray obscuration in BLAGN tends to coincide with optical dust extinction, which is optically thinner than that in the NLAGN and can be partial-covering to the BLR.

\subsection{A Geometrical Torus Model}
\label{sec:torusmodel}
\begin{figure}[htbp]
\epsscale{1}
\plotone{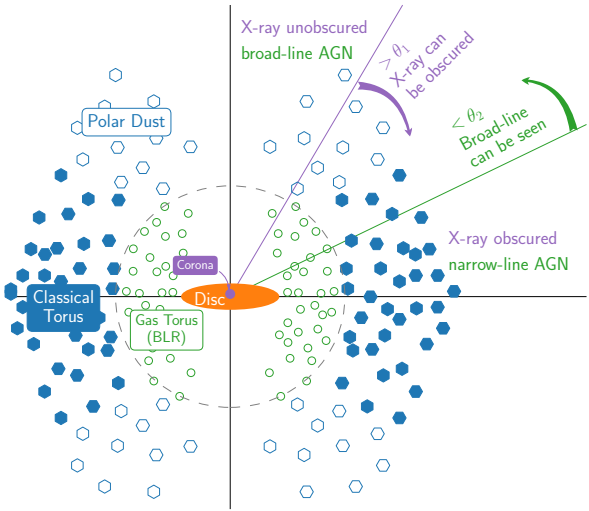}
\caption{Cartoon of the circumnuclear environment of AGN (not to scale). The dashed-line circle indicates the sublimation radius.
Circular clouds inside the sublimation radius are dust free, and hexagon clouds outside the radius are dusty.
The empty clouds (BLR and polar dust) are optically thin to the optical emission, and the filled ones are optically thick.
Above an inclination angle $\theta_1$, the overall column density of the gas and dust becomes sufficient to obscure the X-ray emission from the corona (purple).
Below an inclination angle $\theta_2$, the BLR could leak through the optically thick dusty torus.
}
\label{fig:torus_model}
\end{figure}

We summarize the properties of the obscuring material in BLAGN as follows.
\begin{enumerate}
\item The X-ray absorber in BLAGN2 is similar as in NLAGN but has an optically thinner dust component.
\item The accretion disc in BLAGN2 suffers more dust extinction than in BLAGN1 but, of course, not as thick as in NLAGN.
\item In dust extincted BLAGN, the BLR could also be dust extincted, similarly as in NLAGN, but by a partial-covering and optically thinner absorber.
\end{enumerate}
Clearly, from both the X-ray and optical point of view, BLAGN2 take an intermediate place between BLAGN1 and NLAGN.
As described below, such an intermediate type can be naturally explained by a multi-component, clumpy torus model.

It was known from the very beginning that torus is most likely clumpy \citep{Krolik1988}.
Observationally, clumpy torus models are supported by the fast X-ray absorption variability \citep[e.g.,][]{Markowitz2014,Kaastra2014,Marinucci2016} and the isotropy level of IR emission \citep[e.g.,][]{Levenson2009,Ramos2011}. They have also been successful in explaining the SED and IR spectroscopy of AGN \citep{Ramos2009,Honig2010,Mor2009,Alonso2011,Lira2013}.
We illustrate a clumpy torus in Fig.~\ref{fig:torus_model}.
The dashed-line circle around the central engine (disc+corona) indicates the sublimation radius.
The classical dusty torus (blue filled hexagons) is located outside this radius.
However, if defined as the X-ray absorber, the torus should extend into this radius and have a gaseous part (green empty circles).
This part might contribute in or be identical to the BLR \citep{Goad2012,Davies2015}.
In some local AGNs, IR interferometric observations find an additional dust component in the polar region (blue empty hexagons) beside the classical torus.
This component is optically thin but emits efficiently in MIR, possibly due to an outflowing dusty wind or due to dust in the NLR \citep{Honig2012,Honig2013,Tristram2014,Lopez2014,Lopez2016,Asmus2016}.
All the three components -- the classical dusty torus, the gaseous inner torus (or BLR), and the polar dust -- contribute in the X-ray obscuration. However, the last one can be negligible in terms of column density compared with the other two.

Under the clumpy torus model, the incidence of obscuration along the line of sight is probabilistic in nature.
However, considering the geometric structure of three obscuring components as shown in Fig.~\ref{fig:torus_model}, the obscuring possibility clearly increases with the inclination angle.
We can imagine an inclination angle $\theta_1$, above which the corona becomes X-ray obscured, and an inclination angle $\theta_2$, below which the BLR can be seen.
The typical BLAGN1 and NLAGN are seen at low inclination angles $<\theta_1$ and at high inclination angles $>\theta_2$, respectively.
Among the three components, only the equatorial dusty clouds (blue filled hexagons) could efficiently block the BLR; the optically-thin polar dust (blue empty hexagons) might only reduce the broad line flux moderately.
Also considering that the BLR is an extended region with a much larger scale than that of the corona, it is natural that $\theta_2>\theta_1$.
The intermediate inclination region between $\theta_1$ and $\theta_2$ is where the BLAGN2 reside.
As illustrated in Fig.~\ref{fig:torus_model}, the optical extinction of the BLAGN2 can be attributed to either the optically-thin polar dust or the dusty clumps of the classical torus.
In the latter case, the dust extinction can be optically thin in terms of a small line-of-sight number of clumps.
Future multi-band spectroscopic surveys might allow us to constrain the model in quantity by means of the fraction of BLAGN2 among the entire AGN population.

\subsection{Physical driver of the obscuration incidence}
\label{sec:physicaldriver}
We have shown in \S~\ref{sec:comparison} that BLAGN2 have higher single-epoch \mbh{} and thus lower \lbd{} than BLAGN1.
It is possible that the main physical driver of whether the X-ray emission of a BLAGN is obscured is the \lbd{}, which regulates the covering factor of the X-ray absorber by means of radiation pressure, as pointed out by \citet{Ricci2017} for the X-ray obscuration in NLAGN.
However, in the framework of the torus model described above, we can alternatively attribute all the differences between BLAGN1 and BLAGN2 to an inclination effect without invoking the \lbd{}--driven effect.

We notice that the higher \mbh{} of our BLAGN2 is entirely caused by their larger FWHM of broad lines.
It has been shown by plenty of works that broad line FWHM increases with increasing inclination angle, likely because of a disc-like structure of BLR, and the virial $f$ factor should decrease with increasing inclination angle \citep{Wills1986,Risaliti2011,Pancoast2014,Shen2014,Bisogni2017,MejiaRestrepo2018}.
As discussed in \S~\ref{sec:torusmodel}, BLAGN2 can be explained as BLAGN with high inclination angles.
As a consequence, the larger \mbh{} of BLAGN2 could just result from the failure to consider the inclination effect in the \mbh{} calculation.

Obviously, there is a degeneracy between the $\lambda_{Edd}$-driven effect and inclination effect in explaining the incidence of obscuration. These two explanations are not mutually exclusive. However, we remark that, in the framework of our multi-component, clumpy torus model, the inclination effect simultaneously explains all the findings of this work, including the existence of BLAGN2, the correlation between \mbh{} and X-ray obscuration, the correlation between X-ray obscuration and optical extinction, and the correlation between relative broad line strength and optical extinction.
Meanwhile, attributing the larger \mbh{} of BLAGN2 to larger inclination angles, our model naturally explains why we find a correlation between the \mbh{} and X-ray obscuration but not between the \mbh{} and optical extinction.
As illustrated in Fig.~\ref{fig:torus_model}, the X-ray absorber, composed of the BLR (inner gaseous torus) and the classical dusty torus, has a toroidal or disc-like shape. It is strongly anisotropic even within the inclination range of BLAGN ($<\theta_2$), presenting a steep gradient of the average X-ray obscuring column density as a function of inclination angle.
However, the optical absorber of BLAGN, composed of the polar dust and the low-inclination, low-density part of the classical dusty torus, is more evenly distributed within the BLAGN inclination range. The anisotropy of the optical absorber could become prominent only when it comes into the regime of NLAGN.

\acknowledgments
JXW acknowledges support from 973 program 2015CB857005, NSFC-11421303, and CAS Strategic Priority Research Program XDB09000000.
Y.S. acknowledges support from an Alfred P. Sloan Research Fellowship and NSF grant AST-1715579.

Funding for SDSS-III has been provided by the Alfred P. Sloan Foundation, the Participating Institutions, the National Science Foundation, and the U.S. Department of Energy Office of Science. The SDSS-III web site is http://www.sdss3.org/.

SDSS-III is managed by the Astrophysical Research Consortium for the Participating Institutions of the SDSS-III Collaboration including the University of Arizona, the Brazilian Participation Group, Brookhaven National Laboratory, Carnegie Mellon University, University of Florida, the French Participation Group, the German Participation Group, Harvard University, the Instituto de Astrofisica de Canarias, the Michigan State/Notre Dame/JINA Participation Group, Johns Hopkins University, Lawrence Berkeley National Laboratory, Max Planck Institute for Astrophysics, Max Planck Institute for Extraterrestrial Physics, New Mexico State University, New York University, Ohio State University, Pennsylvania State University, University of Portsmouth, Princeton University, the Spanish Participation Group, University of Tokyo, University of Utah, Vanderbilt University, University of Virginia, University of Washington, and Yale University.

\end{CJK*}
\bibliography{typeIobs}
\appendix
\section{Highly obscured sources}
\label{append:highlyobscured}
\begin{figure}[htbp]
\epsscale{0.6}
\plotone{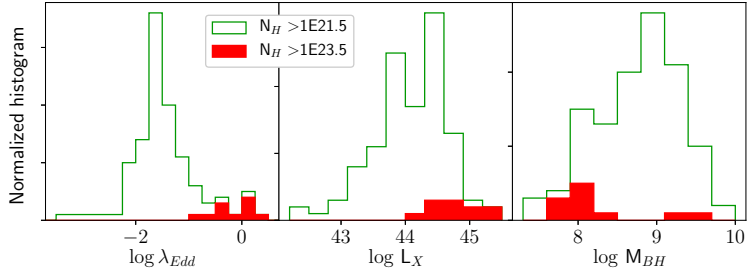}
\caption{
The \lbd{}, \mbh{}, and $L_X$ distributions of the BLAGN2. The ones with \N{H}$>10^{23.5}$~cm$^{-2}$ are filled with red color.
}
\label{fig:obs_23.5}
\end{figure}
Before the sample re-selection, we exclude 11 highly-obscured BLAGN2 with \N{H}$>10^{23.5}$~cm$^{-2}$.
Such sources have very different properties from the other BLAGN2, as shown in Fig.~\ref{fig:obs_23.5}.
They have significantly higher $L_X$ but not accordingly higher \mbh{}.
Conversely, most of them have much lower \mbh{} than the other BLAGN2.
As a consequence, they appear as a high-end tail of the \lbd{} distribution.
Their high \lbd{} might be a result of the effective Eddington limit, which increases with \N{H} (Fig.~\ref{fig:NH_Edd_forbid}), in combination with with the X-ray flux limit of the sample.
However, in such highly-obscured cases, the X-ray absorption correction on the basis of the XMM-Newton spectra (mostly below 10~keV) is highly model-dependent.
Their $L_X$ are less reliable and can be overestimated.
It is also possible that their \mbh{} are underestimated because of dust extinction of their optical emission.

\section{Outflowing Sources}
\label{append:outflow}
As shown in Fig.~\ref{fig:NH_Edd_forbid}, there are three BLAGN2 above the 2-times-corrected effective Eddington limit of a low-dust-fraction absorber (the red dashed line, see \S~\ref{sec:effedd} for details).
Here we also consider the BLAGN2 which is below but the nearest to this limit.
The ID \citep{Liu2016} and redshifts of these four sources are 
N\_89\_36 at $z=1.00$,
N\_66\_6 at $z=0.73$,
N\_64\_36 at $z = 0.49$,
 and N\_160\_16 at $z = 2.34$, with \lbd{} from low to high.
In such cases, unless the X-ray absorber is completely dust free or very far away from the black hole, it should be swept out by the radiation pressure \citep{Fabian2009}. In other words, outflow is expected.
Therefore, we check whether their optical spectra show signs of outflow.

Firstly, all of them have red ($\alpha' > 0$) optical continua (see Fig.~\ref{fig:NH_Edd_forbid}), similar to other sources with outflow, which often show dust extinction \citep[e.g.,][]{Brusa2015,Zakamska2016}.
Meanwhile, almost all the other sources in the dust blow-out region (blue solid line in Fig.~\ref{fig:NH_Edd_forbid}) have $\alpha'<0$.

\begin{figure}[htbp]
\epsscale{0.6}
\plotone{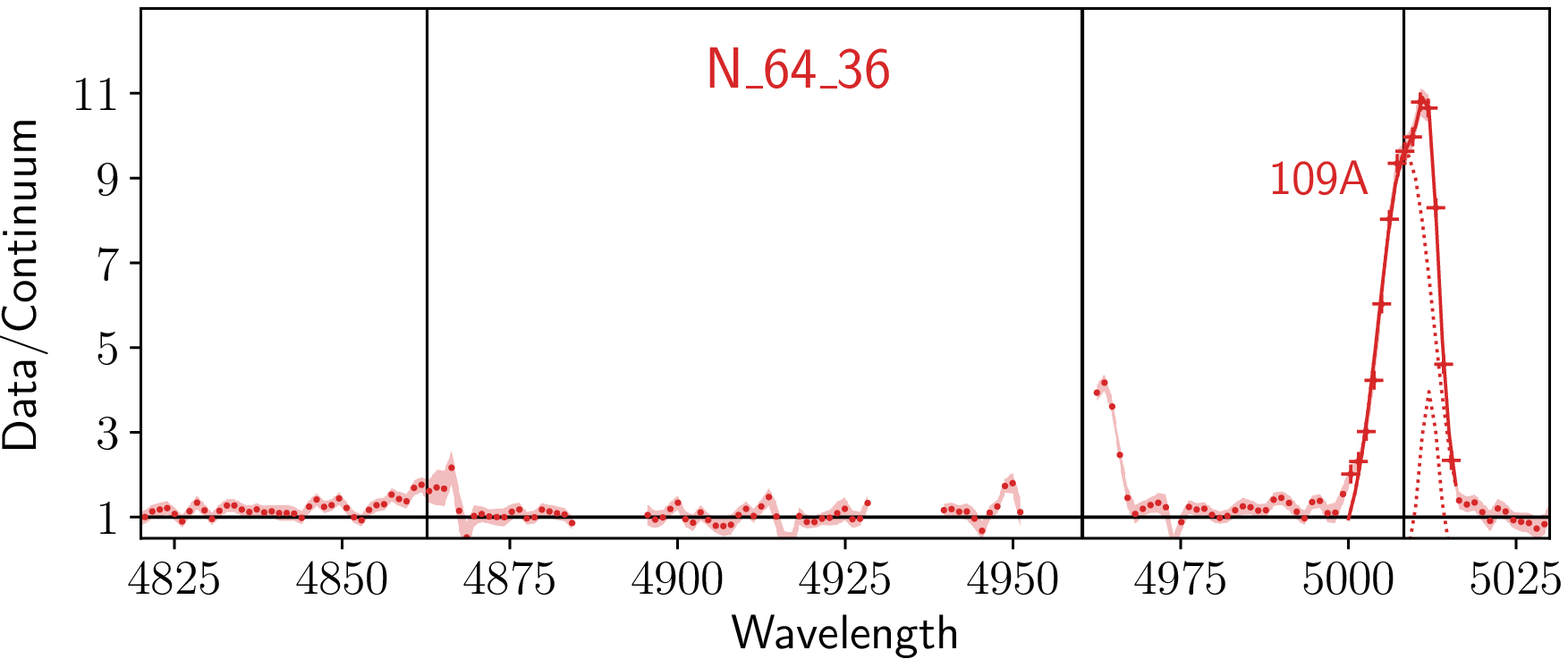}
\plotone{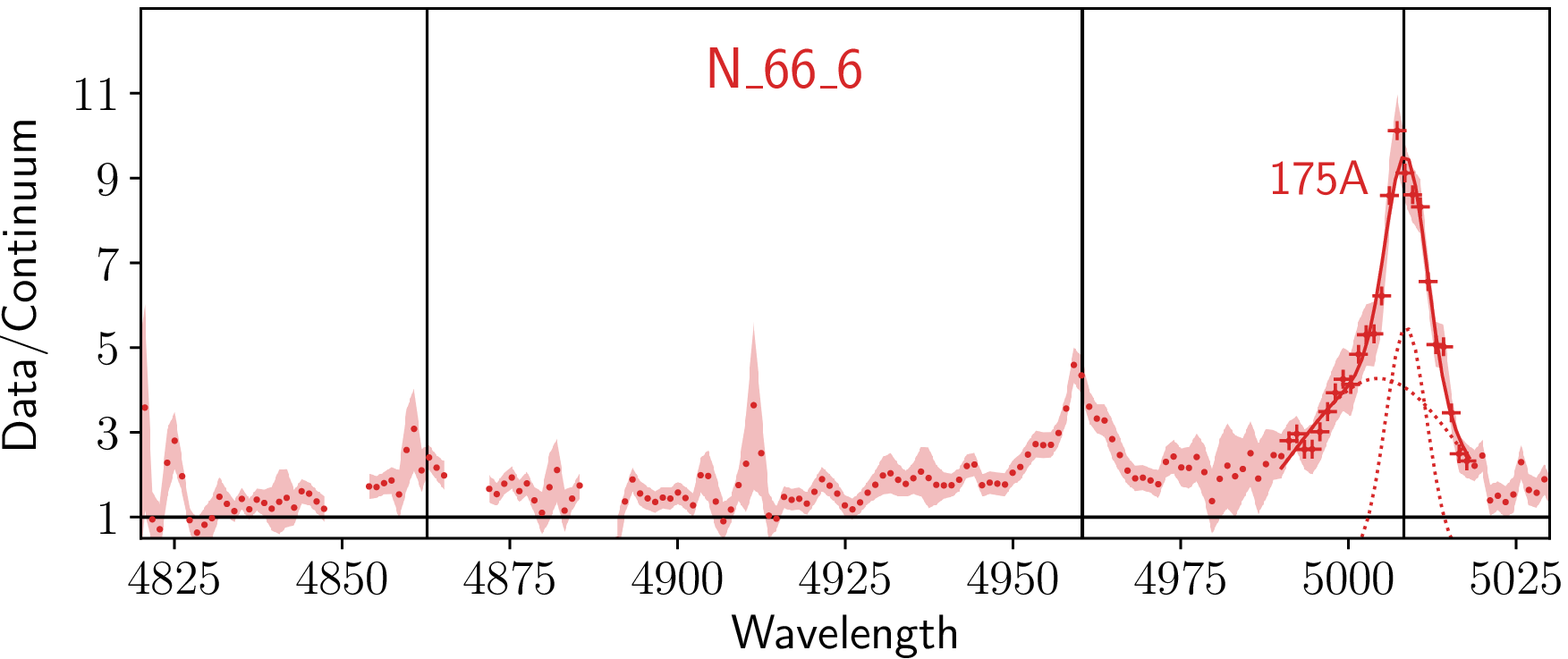}
\caption{The ratio of data to local power-law continuum for the source N\_64\_36 and N\_66\_6. The same set of markers and lines are used as in Fig.~\ref{fig:alllines}.
The EW of the \Oline{III} 5007 line measured by the double-gaussian fit are marked in the figure.
Noise dominated sections of the spectrum are eliminated.
}
\label{fig:outflow}
\end{figure}

Among the four sources, the two low-redshift ones could sample the H$\beta$ and \Oline{III} wavelength range in their SDSS spectra.
We plot their spectra in this range in Fig.~\ref{fig:outflow}.
We note that there are a lot of examples of outflowing sources found with no or very weak H$\beta$ line \citep[e.g.,][]{Brusa2015,Kakkad2016}.
These two sources also show very weak H$\beta$ lines which are almost absent.

We fit the \Oline{III} 5007 lines of the two sources with a double-gaussian profile.
As shown in Fig.~\ref{fig:outflow}, both their \Oline{III} 5007 lines present asymmetric shapes with strong outflowing (blue-shifted) components
\footnote{The asymmetric shape of \Oline{III} is causing a slight underestimation of redshift of N\_64\_36.}
, similar to what is conventionally used to select objects with outflows \citep[e.g.,][]{Harrison2014,Perna2017}.
We argue that, in such cases, it might be the outflowing polar dust that reddens the optical continua and weakens the broad H$\beta$ line.

\section{Composite Spectrum of BLAGN}
\label{append:spec}
\begin{figure}[htbp]
\begin{center}
\epsscale{0.6}
\plotone{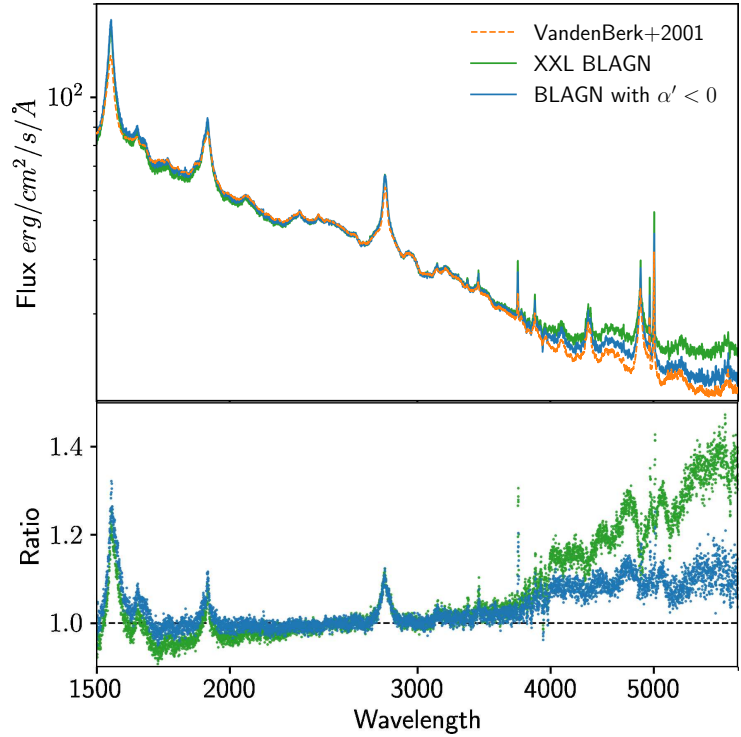}
\caption{
The median composite spectrum generated using method ``A'' (\S~\ref{sec:stackingmethod}) for our BLAGN (sample ``1'', green) and for the blue ($\alpha'<0$) subsample (blue).
The $68\%$ confidence intervals are very narrow and thus not shown.
For comparison, we plot the composite quasar spectrum of \citet{VandenBerk2001} (orange dashed line).
The spectra are normalized between 3020 and 3100$\angstrom$.
The lower panel shows the ratios of our composite spectra to that of \citet{VandenBerk2001}.
}
\label{fig:spec_blue}
\end{center}
\end{figure}

Although, theoretically, geometric mean spectrum has an advantage of preserving the power-law continuum shape over median spectrum, this advantage is impractical in practice, because the spectra are not always power-laws -- a small fraction of them show a reddening caused by dust extinction (see Fig.~\ref{fig:eachspec}).
The median composite spectrum is of more interest and has been used as a cross-correlation template.
In Fig.~\ref{fig:spec_blue}, we show the median composite spectrum of our BLAGN in comparison with the median composite quasar spectrum obtained by \citet{VandenBerk2001}.

Our composite spectrum shows stronger emission lines, a flatter power-law below 4000$\angstrom$, and a red excess above 4000$\angstrom$.
All these differences are caused by different sample selections -- \citet{VandenBerk2001} used color-selected quasars but our BLAGN are selected on the basis of X-ray brightness and optical emission lines.
Firstly, our sample tend to select sources with stronger emission lines.
Meanwhile, we could select the BLAGN in spite of moderate dust extinction of the continuum emission from the disc. Such dust reddened sources are responsible for the flatter power-law of our composite spectra.
The red excess above 4000$\angstrom$ corresponds to a stronger stellar component. 
The relative strength of the stellar component is various among different samples because it depends on fiber diameter, redshift, and AGN luminosity (see also the discussion in \citet{Pol2017}).
Excluding the significantly reddened BLAGN with $\alpha'>0$, the composite spectrum of our BLAGN (blue line and points in Fig.~\ref{fig:spec_blue}) is more similar to that of \citet{VandenBerk2001}.

\end{document}